\documentclass{aastex61}
\usepackage{natbib}
\usepackage[version=3]{mhchem} 
\usepackage{graphicx}

\newcommand\aastex{AAS\TeX}

\accepted{for publication in ApJ}

\shorttitle{\aastex\ Dust composition from impactors}
\shortauthors{Pignatale et al.}

\begin{document}

\title{On the impact origin of Phobos and Deimos III:\\ resulting composition from different impactors}

\correspondingauthor{Francesco C. Pignatale}
\email{pignatale@ipgp.fr}

\author[0000-0003-0902-7421]{Francesco C. Pignatale}
\affil{Institut de Physique du Globe de Paris (IPGP), 1 rue Jussieu, 75005, Paris, France}

\author{S\'ebastien Charnoz}
\affil{Institut de Physique du Globe de Paris (IPGP), 1 rue Jussieu, 75005, Paris, France}
\affil{Institut de Physique du Globe/Universit´e Paris Diderot/CEA/CNRS, 75005 Paris, France}
\author{Pascal Rosenblatt}
\affil{Royal Observatory of Belgium, Avenue circulaire 3, B-1180 Uccle, Belgium}
\affil{now at ACRI-ST, 260 route du pin-montard- BP 234- F-06904 Sophia-Antipolis Cedex, France}
 
\author{Ryuki Hyodo}
\affil{Earth-Life Science Institute/Tokyo Institute of Technology, 152-8550 Tokyo, Japan}

\author{Tomoki Nakamura}
\affil{Tohoku University, 980-8578 Miyagi, Japan}

\author{Hidenori Genda}
\affil{Earth-Life Science Institute/Tokyo Institute of Technology, 152-8550 Tokyo, Japan}



\begin{abstract}

The origin of Phobos and Deimos in a giant impact generated disk is gaining larger attention. Although this scenario has been the subject of many studies, an evaluation of the chemical composition of the Mars' moons  in this framework is  missing. The chemical composition of Phobos and Deimos is unconstrained. The large uncertainness about the origin of the mid-infrared features, the lack of absorption bands in the visible and near-infrared spectra, and the effects of secondary processes on the moons' surface make the determination of their composition very difficult from remote sensing data. Simulations suggest a formation of a disk made of gas and melt with their composition linked to the nature of the impactor and Mars. Using thermodynamic equilibrium we investigate the composition of dust (condensates from gas) and solids (from a cooling melt) that result from different types of Mars impactors (Mars-, CI-, CV-, EH-, comet-like). Our calculations show a wide  range of possible chemical compositions and noticeable differences between dust and solids depending on the considered impactors. Assuming  Phobos and Deimos as result of the accretion and mixing of dust and solids, we find that the derived assemblage (dust rich in metallic-iron, sulphides and/or carbon, and quenched solids rich in silicates)  can  be compatible with the observations. The JAXA's MMX (Martian Moons exploration) mission will investigate the physical and chemical properties of the Maroons, especially sampling from Phobos, before returning to Earth. Our results could be then used to disentangle the origin and chemical composition of the pristine body that hit Mars  and suggest  guidelines for helping in the analysis of the returned samples.

\end{abstract}




\section{Introduction} \label{sec:intro}

The history of  formation of the Mars' moons Phobos and Deimos is still an open question. It has been the subject of several studies which point to a capture origin, in-situ or impact generated formation \citep[and references therein]{2011A&ARv..19...44R,2015Icar..252..334C,2016NatGe...9..581R}. Accretion within an impact-generated disk scenario  \citep[]{2011Icar..211.1150C,2016NatGe...9..581R} is gaining more support as it can explains several properties of the Mars' moons such as the mass and the orbital parameters \citep{2016NatGe...9..581R,hesselbrock2017ongoing,2017ApJ...845..125H,2017ApJ...845..125HBIS}.

Phobos has a very peculiar infrared spectra. Although mid-infrared (MIDIR) show different features, the visibile (VIS) and near-infrared (NIR) spectra are characterized by a lack of absorption features \citep[]{1999JGR...104.9069M,2011P&SS...59.1308G,2011A&ARv..19...44R,2015aste.book..451M}. \citet{1999JGR...104.9069M} isolated two main regions named ``red" and ``blue"  on the Phobos' surface that have different spectral characteristics which can be best matched by  D- and T-type asteroids respectively \citep[]{1991JGR....96.5925M,1999JGR...104.9069M,2002Icar..156...64R}. \citet{2011P&SS...59.1308G} presented a detailed investigation on the possible chemistry of Phobos' surface. They found that the ``blue" region  can be fitted with a phyllosilicates-rich material, while the ``red" region has a best fit when tectosilicates, mainly feldspar, are included in the model.  Moreover they found that no class of  chondritic material  can match the observed spectra. Nevertheless, they pointed out that different more complex mixtures of dust could be able to  reproduce the observed trends.

The featureless VIS-NIR spectra are often associated with a strong  space weathering \citep{1999JGR...104.9069M,2011A&ARv..19...44R}. However, \citet{2011P&SS...59.1308G} following the spectroscopical studies of \citet{1981JGR....86.7967S}, \citet{1989JGR....94.9192S},\citet{1990JGR....95.8323C}, \citet{1990JGR....95..281C}, \citet{1990Icar...84..315C}, \citet{1990Icar...86..383C}, \citet{burns1993mineralogical}, \citet{2007M&PS...42..235K} list a series of possible mechanisms that can reduce the strength of the spectra and match the observation: i) as the 1-2$\mu$m feature arise from iron-bearing material such as pyroxene and olivine, the absence of those compounds may reduce the spectra; ii) a mixture of opaque material such as metallic iron, iron oxides and amorphous carbon mixed with olivine and pyroxene can reduce dramatically the VIS/NIR bands; iii) solids which results from quenching from the liquids state may have their reflectance properties reduced as they lack of perfect crystalline structure; iv) the reflectance of fine-grain materials decreases as the size of the grains decreases.

\citet{2017ApJ...845..125H} presented  detailed Smoothed Particle Hydrodynamics (SPH) simulations in which they determined the dynamical, physical and thermodynamical properties of an impact-generated disk. They found that the material that populate the disk is initially a mixture of gas ($\sim 5$\%) and melts ($\sim 95$\%). These information together with the  Martian composition and hypothesis on the impactors, can be used for modelling the building blocks of Phobos  and Deimos.

In this work we present a study of the bulk composition of the Mars's moons following the giant-impact scenario. Our aim is to provide more clues on the origin of the moons, their chemical composition, infrared spectra, and the nature of the impactor itself.

Furthermore, the JAXA's MMX\footnote{http://mmx.isas.jaxa.jp/en/index.html} mission plans to observe Phobos and Deimos in detail, and return samples (at least 10g) from the surface of Phobos. Our results could be then used as  guidelines for helping in their analysis and interpretation.

Starting from different initial compositions of the impactor (from mars-like to chondritic-like), we compute thermodynamic equilibrium \citep{DeHoff1993} to solve for stable phases that may condense from the gas in the impact-generated disk. Additionally, we compute the composition of the cooling melt to investigate how it will eventually differs from condensates. The resulting condensates and solidified melt are then taken as proxies for the building block of Phobos and Deimos and further discussion are made.

In this work we will mainly focus on Phobos, as more observation are available and as it will be the main sampling target of the JAXA’s MMX mission. Nevertheless, the formation of Deimos follows the same proposed scenario.

The paper is structured as follow: in section~\ref{methods} we describe the techniques and the model we use in our calculations. In section~\ref{results} we present our results that will be discussed in section~\ref{discussion}. Conclusions are summarized in section~\ref{conclusions}.

\section{Model and methods}
\label{methods}

\citet{2017ApJ...845..125H} calculated that the temperature in the Mars' moons forming region of the disk reaches $T\sim2000$~K just after the impact. The value of $P\sim10^{-4}$~bar is chosen as our fiducial pressure  as it is, for the given temperature, the average  saturation pressure for several mixtures  calculated in \citet{2013ApJ...767L..12V} and the average pressure in the disk profile in \citet{2016ApJ...828..109R} and \citet{2017ApJ...845..125H} where gas and melt coexist. Under these conditions  the material in the disk that comes from Mars and from the impactor will result in a mixture composed of gas and melt \citep{2017ApJ...845..125H}.

\begin{figure}
\center
{\includegraphics[width=0.5\columnwidth]{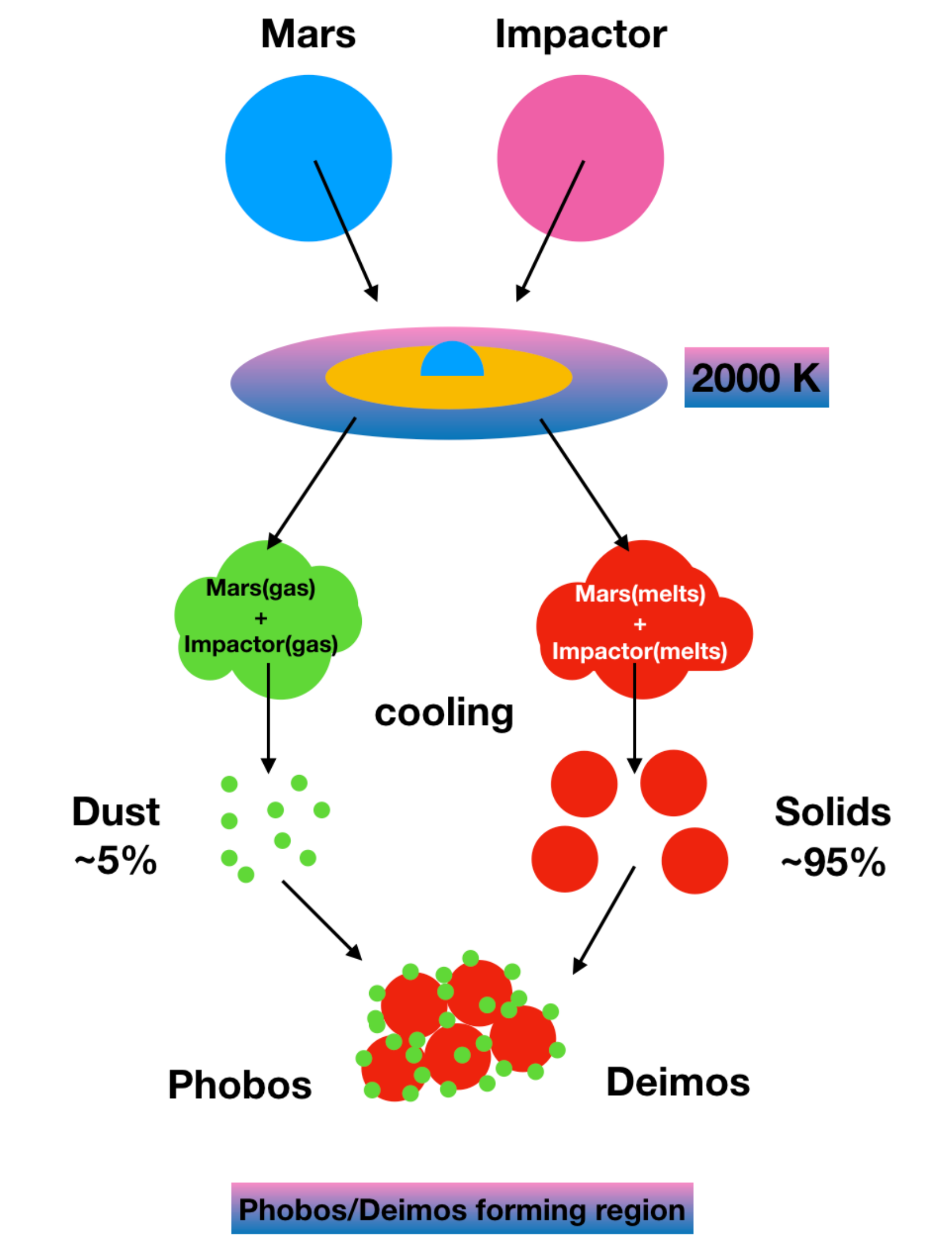}}
\caption{This cartoon describes the considered scenario. After the impact, part of Mars  material will be ejected out at high temperature and will vaporize into gas as well as part of the impactor. The gas mixture will then condense into dust. On the other hand, the not vaporized material from Mars and the impactor will form a melt and then solidify.  Phobos and Deimos will be the result of the accretion of  these two components. The yellow region represents the part of disk within the Roche limit \citep{2017ApJ...845..125H}.}
\label{fig1}
\end{figure}

\citet{2017ApJ...845..125H} showed that the building blocks of Phobos and Deimos would be composed of a mixture of about half-martian material and half-impactor material. We, thus, assume that the gas is made of a well mixed two-components: the gas that is released by heating up Mars-material plus the gas that is released by heating up impactor-material. We then assume that the melt is a mixture of the not vaporized material from the two bodies \citep{2017ApJ...845..125H}.  Figure~\ref{fig1} shows a cartoon of the proposed model. 

As the disk cools down, the gas will eventually re-condense and the melt will solidify. In this work, we define, for ease of understanding, {\it dust} as the condensates from the gas phase and {\it solids} as the material that result from the solidification of the melt.

In order to determine the composition of the dust that will condense from the gas phase we assume thermodynamic equilibrium \citep{DeHoff1993}: at constant temperature and pressure, the stability of a system is determined by its Gibbs free energy, and, in fact, by the composition which minimizes the potential of the system. Although it is an approximation, thermodynamic equilibrium  is a powerful tool to understand the evolution of the chemical composition of complex systems. This technique has been extensively used  in the study of the chemistry of gas and dust in several astrophysical environments: from the Solar Nebula, meteorites and protoplanetary disks \citep{1979Icar...40..446L,Yoneda1995,2003ApJ...591.1220L,2006mess.book..253E,2016MNRAS.457.1359P} to stars dusty envelopes \citep{1999A&A...347..594G,1997AIPC..402..391L,2001GeCoA..65..469E} and exoplanets composition \citep{2010ApJ...715.1050B}. 

To compute the thermodynamic equilibrium we use the HSC software package (version 8) \citep{roine2002outokumpu}, which includes the Gibbs free energy minimisation method of \citet{White1958}. Thermodynamic data for each compound are taken from the database provided by HSC \citep[and references therein]{roine2002outokumpu}. HSC has been widely used in material science and it has been already tested in astrophysics showing very good reliability in predicting the composition of different systems \citep{2005Icar..175....1P,2011MNRAS.414.2386P,2010ApJ...715.1050B,2012ApJ...759L..40M}.

To calculate the  composition of the solids from the cooling melts we use the normative mineralogy (CIPW-norm) \citep{10.2307/30060535} and the work of \citet{2016ApJ...828..109R} as benchmark.   CIPW-norm is one of the most used technique to determine, in a first approximation, the equilibrium composition of a multicomponent  melt \citep{10.2307/30060535}.

\citep{2017ApJ...845..125H} showed that the melt phase of Mars and the impactor will likely never completely equilibrate between each other. Mars-only and impactor-only melt with different degrees of equilibration in between are indeed expected. Nevertheless, calculating the resulting compositions of a equilibrated melt represents a first interesting add-on to investigate the differences that condensation and solidification would bring to the final Phobos bulk composition. Moreover, our model suggests that the MMX may be confronted with two distinguishable family of material, the $\it{dust}$ and the $\it{solids}$. As a consequence, this investigation can bring further information and clues that can be used in the MMX samples analysis.

During the planet formation, Mars and the other inner rocky planets experienced impacts with other bodies. The impact histories strongly depend on the timing and location of the planets \citep{2017E&PSL.468...85B,2017Icar..297..134R,2017GeoRL..44.5978B,bottke2017post}. The nature of the impactors is unconstrained as the dynamical interactions of Jupiter and Saturn with the surrounding minor bodies may have scattered and delivered in the inner Solar System material of different nature and of chondritic origin \citep{2017Icar..297..134R,2017GeoRL..44.5978B,bottke2017post}. Our aim is to determine the changes that different impactors would bring in the chemical composition of Phobos, and if these differences can be traceable. In order to keep our selection as chemically heterogeneous as possible we, thus, consider the following types of impactor: Mars$_{type}$, CV$_{type}$, EH$_{type}$, CI$_{type}$, comet$_{type}$. As a proxy of Mars composition we take the  Bulk-Silicates-Mars (BSM) from \citet{2013ApJ...767L..12V}. Compositions for the EH, CV and CI chondrites are taken from \citet{1988RSPTA.325..535W}. Elemental composition for the comet is taken from \citet{Huebner1997}. Table~\ref{table1} shows the elemental distribution for all considered impactors. In order to help to understand the differences between the impactors we also report several elemental ratios such as the Mg/Si, Fe/O, C/O ratios. These ratios play an important role in determining the resulting chemical composition of a mixture. This will be discussed in section~\ref{discussion}.  

We also report values from the Sun photosphere\footnote{We take  the elemental abundances from \citet{2009ARA&A..47..481A} for the given set of elements. Please note that \ce{He} is not included in the system and, as a consequence, the abundance of \ce{H} raises to $\sim$99\% of the total.} as reference \citep{2009ARA&A..47..481A}. Note the H/O ratio of the solar nebula (Sun) and the abundances of other elements  relatives to O and C. Looking at table~\ref{table1}, we can already notice that we will deal with wide different environments. Moreover, the relative abundances between elements clearly indicate that our systems will return chemical distributions that are far from that one predicted for a solar composition.

\begin{table*}													
\centering													
\caption{Elemental abundances (mol\%) for single impactors. Abundances of the solar photosphere (Sun) are also shown.}													
\begin{tabular} {l c c c c c c }													
\hline													
	& 	Mars	&	CV	&	CI	&	EH	&	comet	&	Sun	\\
\hline													
Element	& 	\multicolumn{4}{c}{Abundances (mol\%)}				&		&		\\			
\hline													
Al	& 	1.250E+00	& 	1.356E+00	& 	4.668E-01	& 	7.596E-01	& 	7.000E-02	&	2.82E-04	\\
C	& 	0.000E+00	& 	9.747E-01	& 	3.902E+00	& 	8.427E-01	& 	1.137E+01	&	2.69E-02	\\
Ca	& 	9.300E-01	& 	9.911E-01	& 	3.362E-01	& 	5.367E-01	& 	6.000E-02	&	2.19E-04\\
Fe	& 	5.440E+00	& 	8.797E+00	& 	4.773E+00	& 	1.314E+01	& 	5.200E-01	&	3.16E-03	\\
H	& 	0.000E+00	& 	5.808E+00	& 	2.906E+01	& 	0.000E+00	& 	5.464E+01	&	99.9 	\\
K	& 	4.800E-02	& 	1.658E-02	& 	2.098E-02	& 	5.177E-02	& 	0.000E+00	&	1.07E-05	\\
Mg	& 	1.632E+01	& 	1.247E+01	& 	5.845E+00	& 	1.104E+01	& 	9.900E-01	&	3.98E-03	\\
Na	& 	7.000E-01	& 	3.001E-01	& 	3.121E-01	& 	7.484E-01	& 	1.000E-01	&	1.74E-04	\\
O	& 	5.815E+01	& 	5.157E+01	& 	4.491E+01	& 	4.724E+01	& 	2.834E+01	&	4.89E-02\\
S	& 	0.000E+00	& 	1.434E+00	& 	2.695E+00	& 	4.577E+00	& 	7.100E-01	&	1.32E-03	\\
Si	& 	1.673E+01	& 	1.624E+01	& 	7.656E+00	& 	2.104E+01	& 	1.830E+00	&	3.23E-03	\\
Ti	& 	4.000E-02	& 	4.280E-02	& 	1.285E-02	& 	2.379E-02	& 	0.000E+00	&	8.90E-06	\\
Zn	& 	2.000E-04	& 	3.709E-03	& 	6.988E-03	& 	9.674E-03	& 	0.000E+00	&	3.63E-06	\\
\hline													
Mg/Si	& 	0.98	& 	0.77	& 	0.76	& 	0.52	& 	0.54	&	1.25	\\
Fe/O	& 	0.09	& 	0.17	& 	0.11	& 	0.28	& 	0.02	&	0.06	\\
Fe/Si	&	0.33	&	0.54	&	0.62	&	0.62	&	0.28	&	0.98	\\
(Fe+Si)/O	&	0.38	&	0.49	&	0.28	&	0.72	&	0.08	&	0.13	\\
C/O	& 	0.00	& 	0.02	& 	0.09	& 	0.02	& 	0.40	&	0.54	\\
H/O	& 	0.00	& 	0.11	& 	0.65	& 	0.00	& 	1.93	&	2041	\\
\end{tabular}													
\label{table1}													
\end{table*}	

The 13 considered elements in Table~\ref{table1} can form  $\sim$6800 possible compounds, including complex organics, gas and solids (and excluding liquids). Most of these compounds are not stable at our chosen $T$ and $P$.  We derive our fiducial list of compounds starting from the list reported in  \citet{2013ApJ...767L..12V} and the set of compounds in \citet{2011MNRAS.414.2386P}. Complex organics have been excluded from calculations as their chemistry is driven more by kinetics rather than thermodynamic equilibrium. \ce{C}-graphite is taken as representative of the main carbon condensates together with \ce{Fe3C}, \ce{Fe2C} and \ce{SiC}. Calcium and aluminium refractory species, all main oxides and main silicates (Mg and Fe silicates)  have been taken into account. Sulfides are included as well as water-vapour and water-ice. We report the complete list of considered species in Table~\ref{table2}. The following nomenclature will be used: olivine (forsterite, \ce{Mg2SiO4},  and fayalite, \ce{Fe2SiO4}), pyroxene (enstatite, \ce{MgSiO3}, and  ferrosilite, \ce{FeSiO3}), plagioclase (anorthite, \ce{CaAl2Si2O8} and albite, \ce{NaAlSi3O8}), melilite (gehlenite, \ce{Ca2Al2SiO7}, and akermanite, \ce{Ca2MgSi2O7}), fassaite (Ca-Tschermak, \ce{CaAl2SiO6},  and diopside, \ce{CaMgSi2O6}), spinel (\ce{MgAl2O4} and \ce{FeAl2O4}), magnesiowustite (\ce{MgO} and \ce{FeO}), sulfide (\ce{FeS}, \ce{MgS} and \ce{CaS}), metal (\ce{Fe}, \ce{Al} and \ce{Zn}). Only the endmembers of each  solids solution are considered and no predictions of intermediate compositions are made.

To summarize, we calculate the thermodynamic equilibrium for each of the considered cases in table~\ref{table1} at the given temperature ($T=2000$~K)  and pressure ($P=10^{-4}$~bar).
The resulting gas phase of Mars plus the gas phase of the selected impactor will constitute the gas mixture from which the ${\it dust}$ will condense. The material that is not in the gas phase will form the melt from which the ${\it solids}$ will form. To derive the ${\it dust}$ composition we then proceed  to the computation of the condensation sequence in the interval of temperatures of $150< T\rm(K)<2000$ with a constant pressure of $P=10^{-4}$~bar. To derive the ${\it solids}$ composition we compute the CIPW-norm.

In order to test our thermodynamic model we also run equilibrium calculation using the solar abundances in Table~\ref{table1}  and compare the results with previous calculations available in the literature. Results of the test and a brief discussion are presented in Appendix~\ref{solar}.
\begin{table*}							
\centering							
\caption{Complete list of gas and dust species in the equilibrium calculations.}			
\begin{tabular}	{c}
\hline
Gas \\
\hline
Al(g) Al2O2(g) Al2O3(g) AlO(g) AlO2(g) \\
C(g) Ca(g) CaO(g) CH4(g) CO(g) \\
Fe Fe(g) FeO(g) FeS(g) \\
H(g) H2(g) H2O(g) H2S(g) HS(g) \\
K(g) K2(g) K2O(g) KO(g) Mg(g) MgO(g) \\
Na(g) Na2(g) Na2O(g) NaO(g) \\
O(g) O2(g) OH(g) \\
S(g) Si(g) SiO(g) SiO2(g)  \\
Ti(g) TiO(g) TiO2(g) \\
Zn(g) ZnO(g) \\
\hline
Dust \\
\hline
Al2O3 \\
C \\
Ca2Al2SiO7 Ca2MgSi2O7 Ca2SiO4 CaAl12O19 CaAl2O4 CaAl2Si2O8  \\
CaAl2SiO6 CaAl4O7 CaMgSi2O6  CaO CaS CaSiO3 CaTiSiO5 \\
Fe Fe2C Fe2O3 Fe2SiO4 Fe3C Fe3O4 \\
FeAl2O4 FeO FeSiO3 FeTiO3\\
H2O \\
K K2O K2Si4O9 KAlSi2O6 KAlSi3O8 KAlSiO4\\
Mg2Al4Si5O18 Mg2SiO4 Mg2TiO4 MgAl2O4 MgO MgS MgSiO3 MgTi2O5 MgTiO3\\
Na2O Na2SiO3 NaAlSi3O8 \\
Si SiC SiO2 \\
TiO2 \\
Zn Zn2SiO4 Zn2TiO4 ZnO ZnSiO3\\
\end{tabular}		
\label{table2}							
\end{table*}

\section{Results}
\label{results}

Table~\ref{table3} shows the elemental abundances (mol\%) of the gas mixture that results from equilibrium calculation at  $T=2000$~K and $P=10^{-4}$~bar of Mars plus the considered impactor (Mars+Mars, Mars+CV, Mars+CI, Mars+EH, Mars+comet). These abundances are used as input to compute the condensation sequence. Table~\ref{table4} show  the oxides budget of different melt mixtures in case of complete equilibration between Mars and given impactors. These budgets are used to compute the CIPW-norm.

\begin{table*}											
\centering											
\caption{Elemental abundances (mol\%) of the gas mixture which is released  after the impact assuming $T=2000~K$, $P=10^{-4}$~bar, different types of impactors, and equal contribution between Mars and the considered impactor.}		
\begin{tabular} {l c c c c c}											
\hline											
Gas mixture	& 	+Mars	&	+CV	&	+CI	&	+EH	&	+comet	\\
\hline											
Element	& 	\multicolumn{4}{c}{Abundances (mol\%)} &				\\				
\hline											
Al	& 	8.059E-06	& 	1.382E-05	& 	5.058E-06	& 	2.336E-05	& 	1.412E-05	\\
C	& 	0.000E+00	& 	3.762E+00	& 	5.791E+00	& 	2.060E+00	& 	1.141E+01	\\
Ca	& 	2.348E-05	& 	4.508E-05	& 	9.962E-06	& 	8.897E-05	& 	1.804E-04	\\
Fe	& 	1.974E+01	& 	2.972E+01	& 	6.652E+00	& 	3.117E+01	& 	8.041E-01	\\
H	& 	0.000E+00	& 	2.241E+01	& 	4.314E+01	& 	0.000E+00	& 	5.481E+01	\\
K	& 	3.360E+00	& 	2.489E-01	& 	1.008E-01	& 	2.423E-01	& 	4.808E-02	\\
Mg	& 	3.539E-01	& 	4.512E-01	& 	1.000E-01	& 	8.458E-01	& 	9.982E-01	\\
Na	& 	4.905E+01	& 	3.859E+00	& 	1.502E+00	& 	3.543E+00	& 	8.026E-01	\\
O	& 	2.608E+01	& 	2.501E+01	& 	3.695E+01	& 	2.775E+01	& 	2.858E+01	\\
S	& 	0.000E+00	& 	5.519E+00	& 	4.000E+00	& 	1.120E+01	& 	7.121E-01	\\
Si	& 	1.146E+00	& 	8.663E+00	& 	1.735E+00	& 	2.310E+01	& 	1.825E+00	\\
Ti	& 	2.663E-01	& 	3.370E-01	& 	2.347E-02	& 	6.733E-02	& 	3.844E-03	\\
Zn	& 	1.051E-02	& 	1.505E-02	& 	1.158E-02	& 	2.398E-02	& 	2.007E-04	\\
\hline											
Mg/Si	& 	0.31	& 	0.05	& 	0.06	& 	0.04	& 	0.55	\\
Fe/O	& 	0.76	& 	1.19	& 	0.18	& 	1.12	& 	0.03	\\
Fe/Si	& 	17.23	& 	3.43	& 	3.83	& 	1.35	& 	0.44	\\
(Fe+Si)/O	& 	0.80	& 	1.53	& 	0.23	& 	1.96	& 	0.09	\\
C/O	& 	0.00	& 	0.15	& 	0.16	& 	0.07	& 	0.40	\\
H/O	& 	0.00	& 	0.90	& 	1.17	& 	0.00	& 	1.92	\\
\end{tabular}											
\label{table3}											
\end{table*}											

\begin{table*}											
\centering											
\caption{Oxide composition (wt\%) of the melt that results after the impact assuming $T=2000~K$, $P=10^{-4}$~bar and different types of impactor. Total equilibration between Mars and impactor is also assumed. The Mg/Si and Fe/O ratios are the elemental mole ratios.}			
\begin{tabular} {l c c c c c}											
\hline											
	& 	+Mars	&	+CV	&	+CI	&	+EH	&	+comet	\\
\hline											
\ce{Al2O3}	& 	2.96	& 	3.55	& 	3.06	& 	3.02	& 	3.10	\\
\ce{CO2}	& 	0.00	& 	0.00	& 	0.00	& 	0.00	& 	0.00	\\
\ce{CaO}	& 	2.40	& 	2.87	& 	2.47	& 	2.41	& 	0.28	\\
\ce{FeO}	& 	17.16	& 	12.55	& 	14.42	& 	12.35	& 	17.06	\\
\ce{H2O}	& 	0.00	& 	0.00	& 	0.00	& 	0.00	& 	0.00	\\
\ce{K2O}	& 	0.00	& 	0.00	& 	0.00	& 	0.00	& 	0.00	\\
\ce{MgO}	& 	30.60	& 	30.88	& 	31.15	& 	32.04	& 	31.09	\\
\ce{Na2O}	& 	0.00	& 	0.00	& 	0.00	& 	0.00	& 	0.00	\\
\ce{SiO2}	& 	46.75	& 	49.35	& 	48.80	& 	50.10	& 	45.11	\\
\ce{TiO2}	& 	0.13	& 	0.80	& 	0.10	& 	0.09	& 	3.37	\\
\ce{ZnO}	& 	0.00	& 	0.00	& 	0.00	& 	0.00	& 	0.00	\\
\ce{SO3}	& 	0.00	& 	0.00	& 	0.00	& 	0.00	& 	0.00	\\
\hline											
Mg/Si	& 	0.98	& 	0.93	& 	0.95	& 	0.95	& 	1.03	\\
Fe/O	& 	0.09	& 	0.06	& 	0.07	& 	0.06	& 	0.09	\\											
\end{tabular}											
\label{table4}											
\end{table*}

\subsection{Dust from condensing gas}
\label{condensationsequence}

\begin{figure}
{\includegraphics[width=0.5\columnwidth]{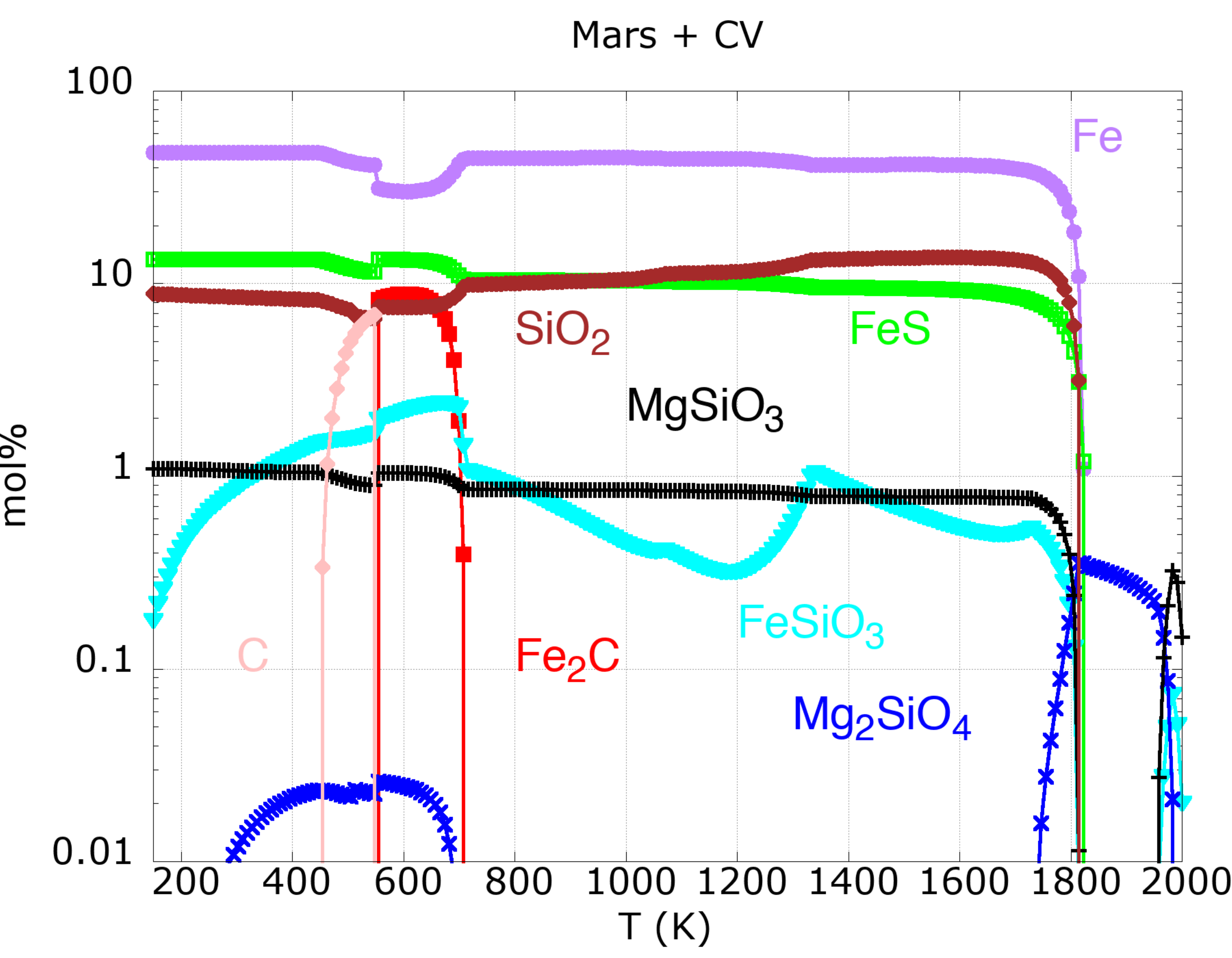}} 
{\includegraphics[width=0.5\columnwidth]{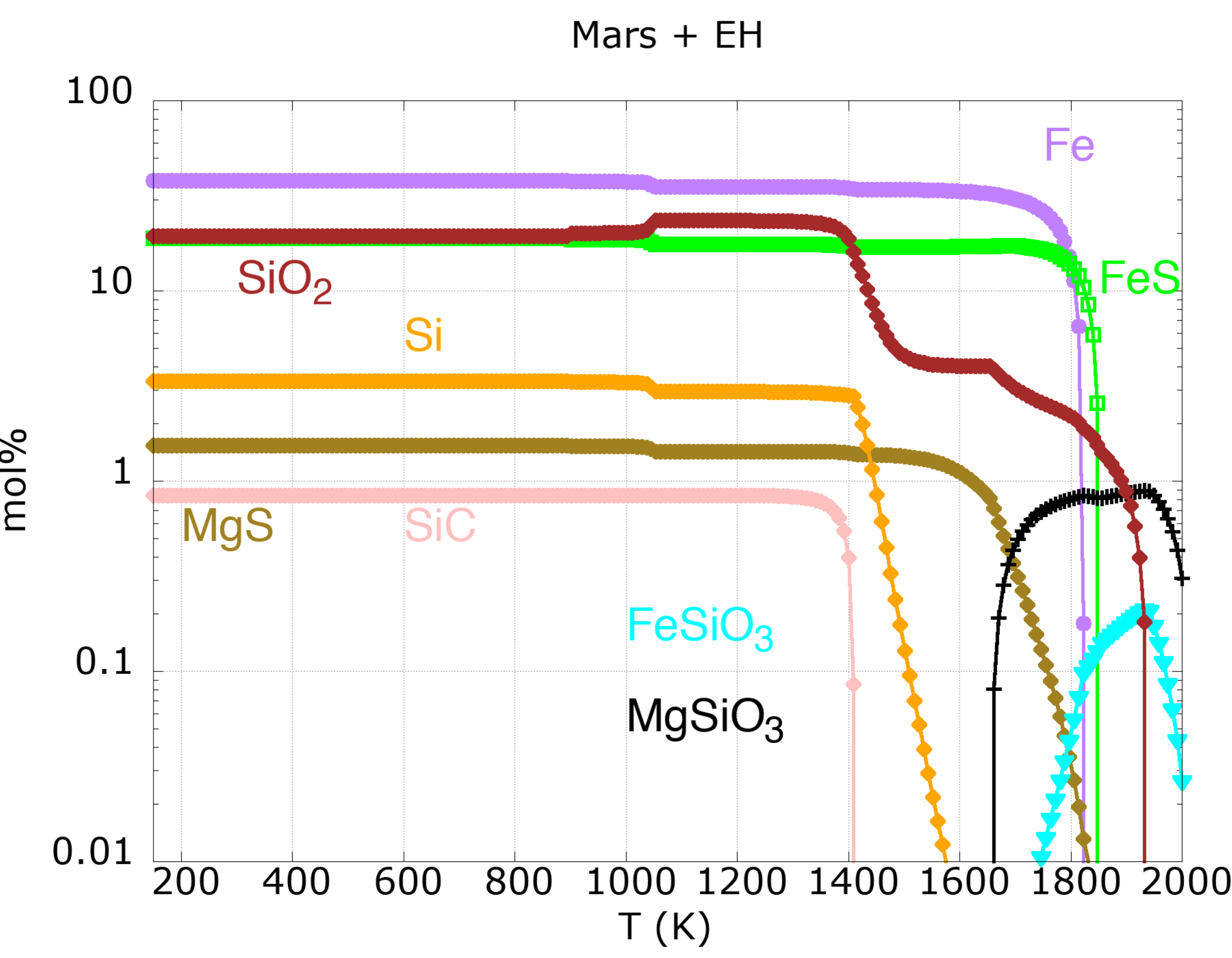}} \\
{\includegraphics[width=0.5\columnwidth]{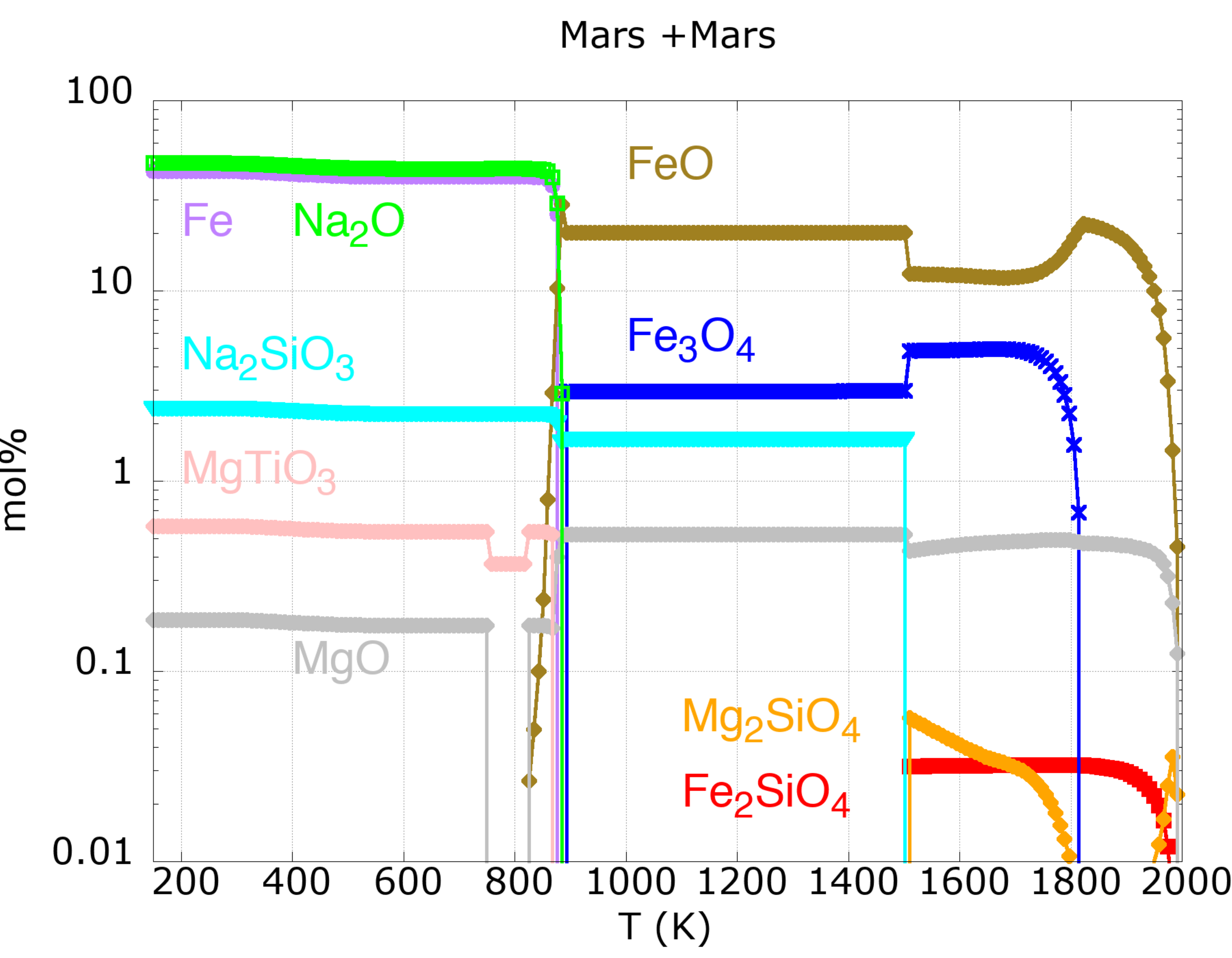}}
{\includegraphics[width=0.5\columnwidth]{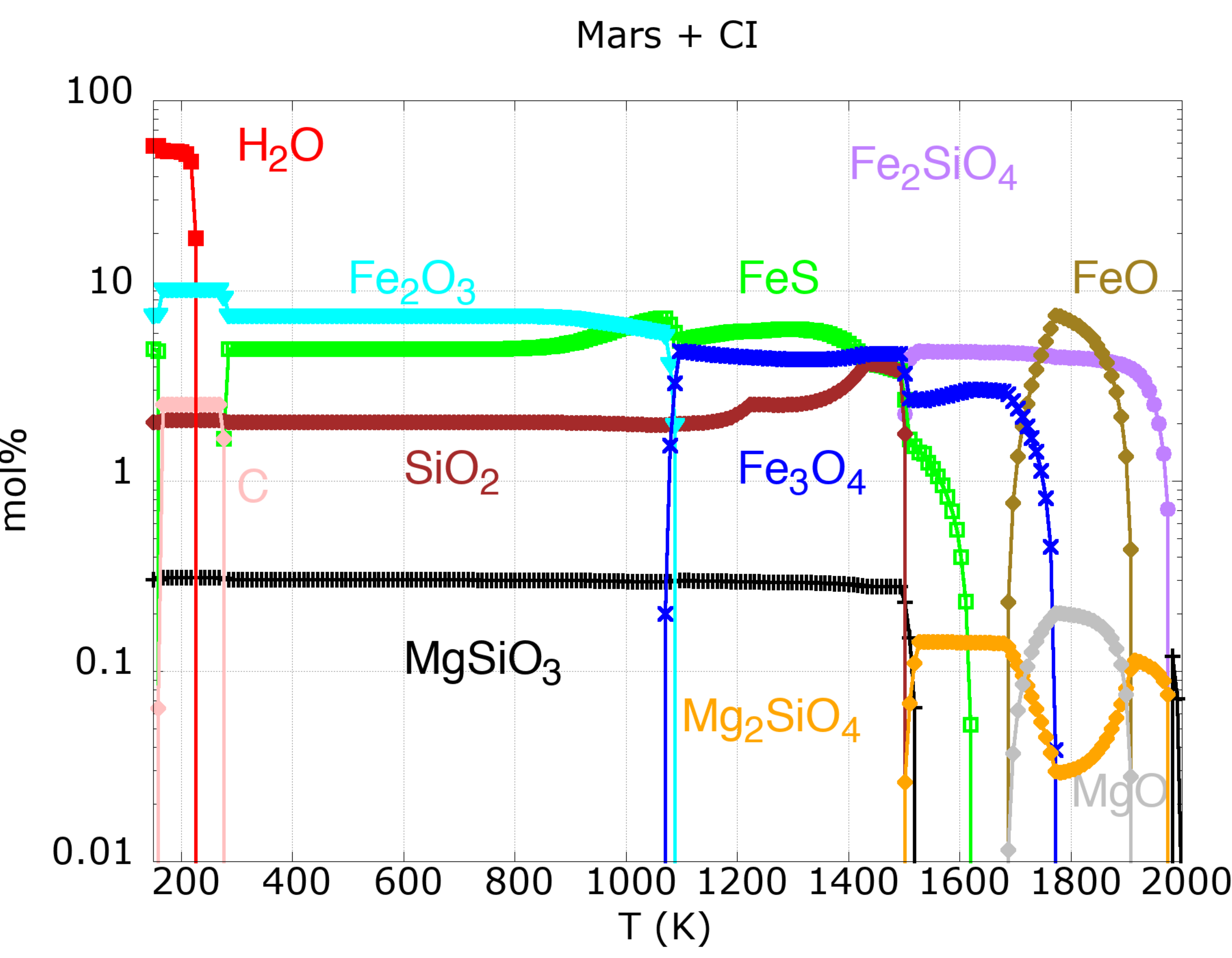}} \\
{\includegraphics[width=0.5\columnwidth]{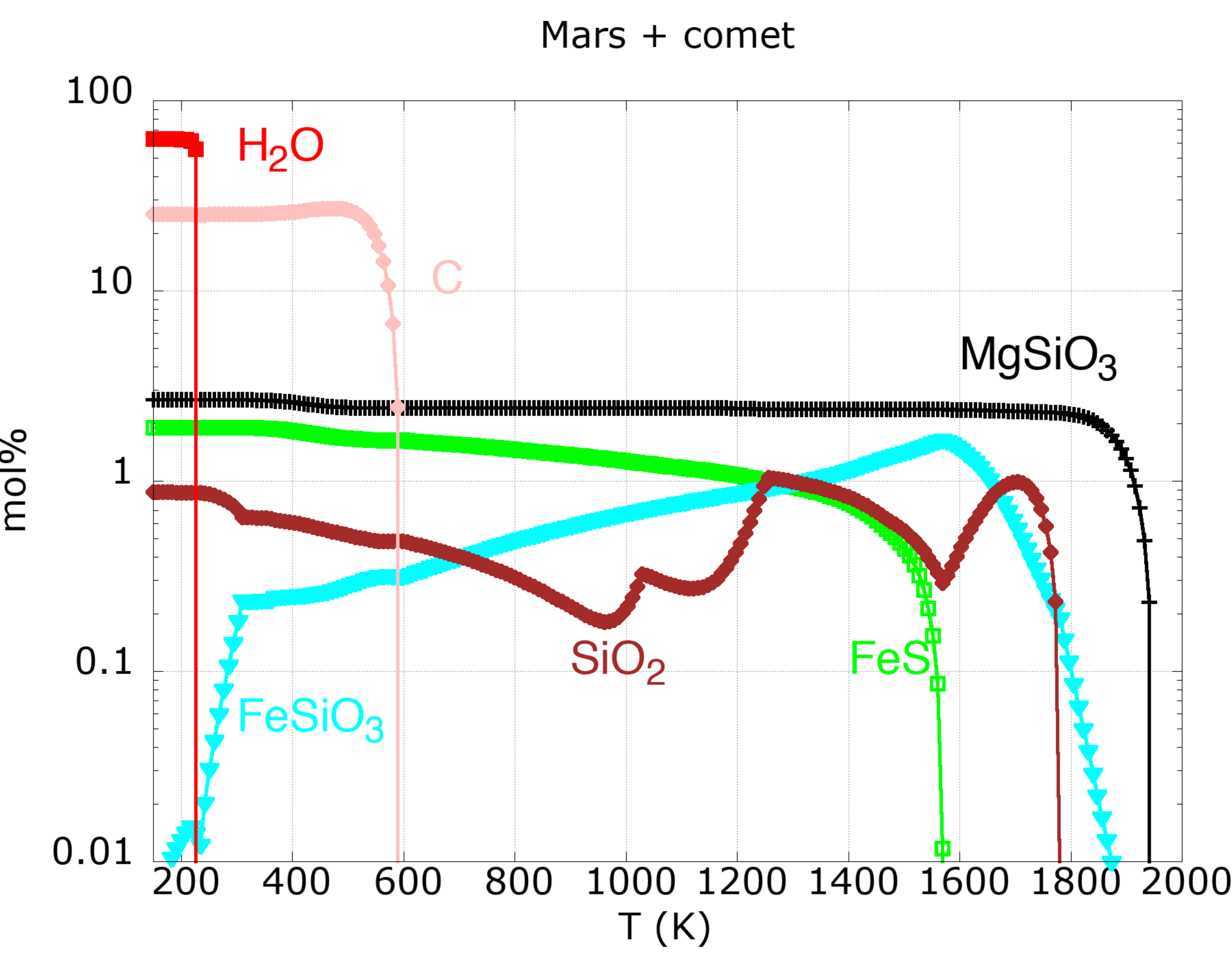}}
\caption{Condensation sequences for major dust species (mol\%) that result from the gas mixtures in Table~\ref{table3}. Note the changes of colours when different compounds are considered.}
\label{fig2}
\end{figure}

Figure~\ref{fig2} shows the  dust distribution for all the considered impactors in mol\% (being 100 gas+dust) as a function of temperature.  From left to right and from top to bottom, the different cases are ordered with decreasing  Fe/O ratio of the initial gas mixture (see Table~\ref{table3}).

Mars+CV impact results in large quantities  of metallic iron, \ce{FeS} and \ce{SiO_2}. Small amount of pyroxene (enstatite (\ce{MgSiO3}) and ferrosilite (\ce{FeSiO3})), $\sim1$~mol\%, is distributed all along the temperature range. At $T~\sim700$~K, we see the appearance of \ce{Fe2C} and C (graphite). Similarly to Mars+CV, the Mars+EH impact shows large quantities of metallic iron, \ce{FeS} and \ce{SiO_2}. Moreover, we do see small percentage of Si, MgS and SiC. Traces of pyroxenes are seen at high temperatures only.

The Mars+Mars impact produces several oxides such as FeO, \ce{Fe3O4}, metallic iron and volatiles such as \ce{Na2O}, \ce{Na2SiO3}. Traces of olivine (forsterite (\ce{Mg2SiO4}) and fayalite (\ce{Fe2SiO4})) are present at high temperature. 

Mars+CI impact returns iron-rich olivine such as fayalite (\ce{Fe2SiO4}), then FeO, \ce{Fe3O4}, \ce{Fe2O3}, \ce{FeS} and smaller amount of \ce{SiO2}. At lower temperature we see the condensation of C and \ce{H2O}. The dust from Mars+comet impact is mainly made of pyroxene, \ce{SiO2} and FeS. Mars+comet impact is that one that produces, as expected, a large amount of water ice together with solid carbon.

\begin{figure}
{\includegraphics[width=0.5\columnwidth]{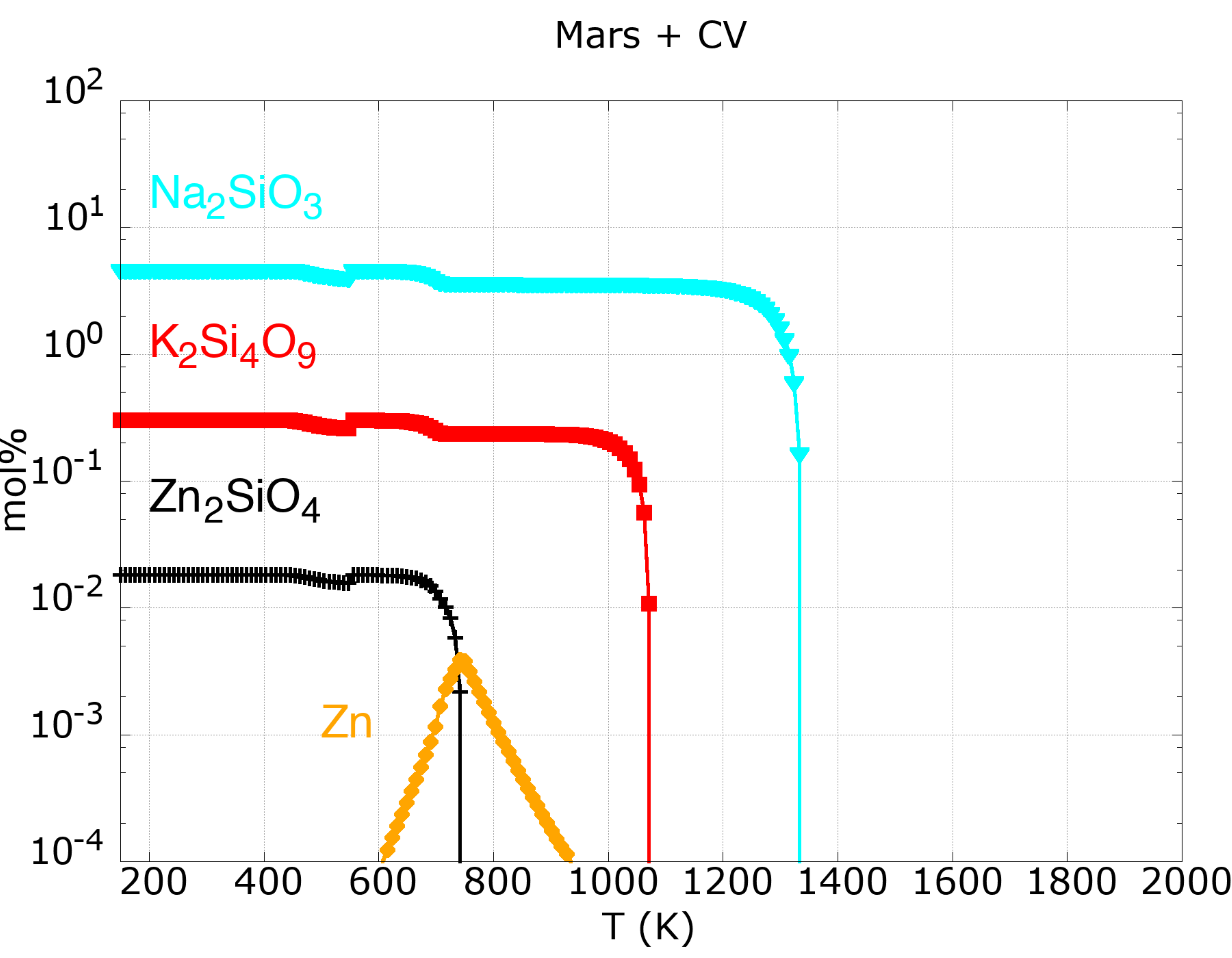}} 
{\includegraphics[width=0.5\columnwidth]{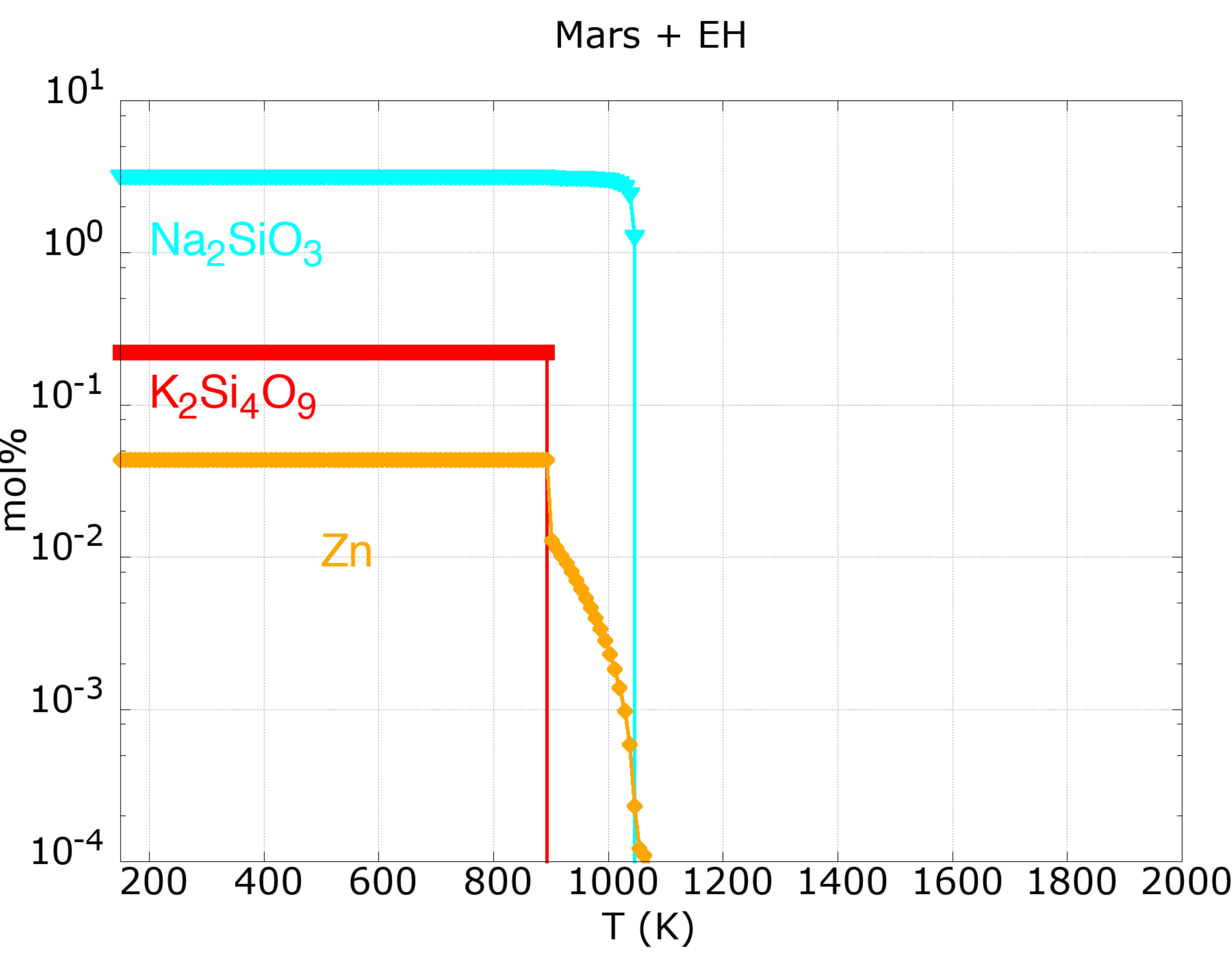}} \\
{\includegraphics[width=0.5\columnwidth]{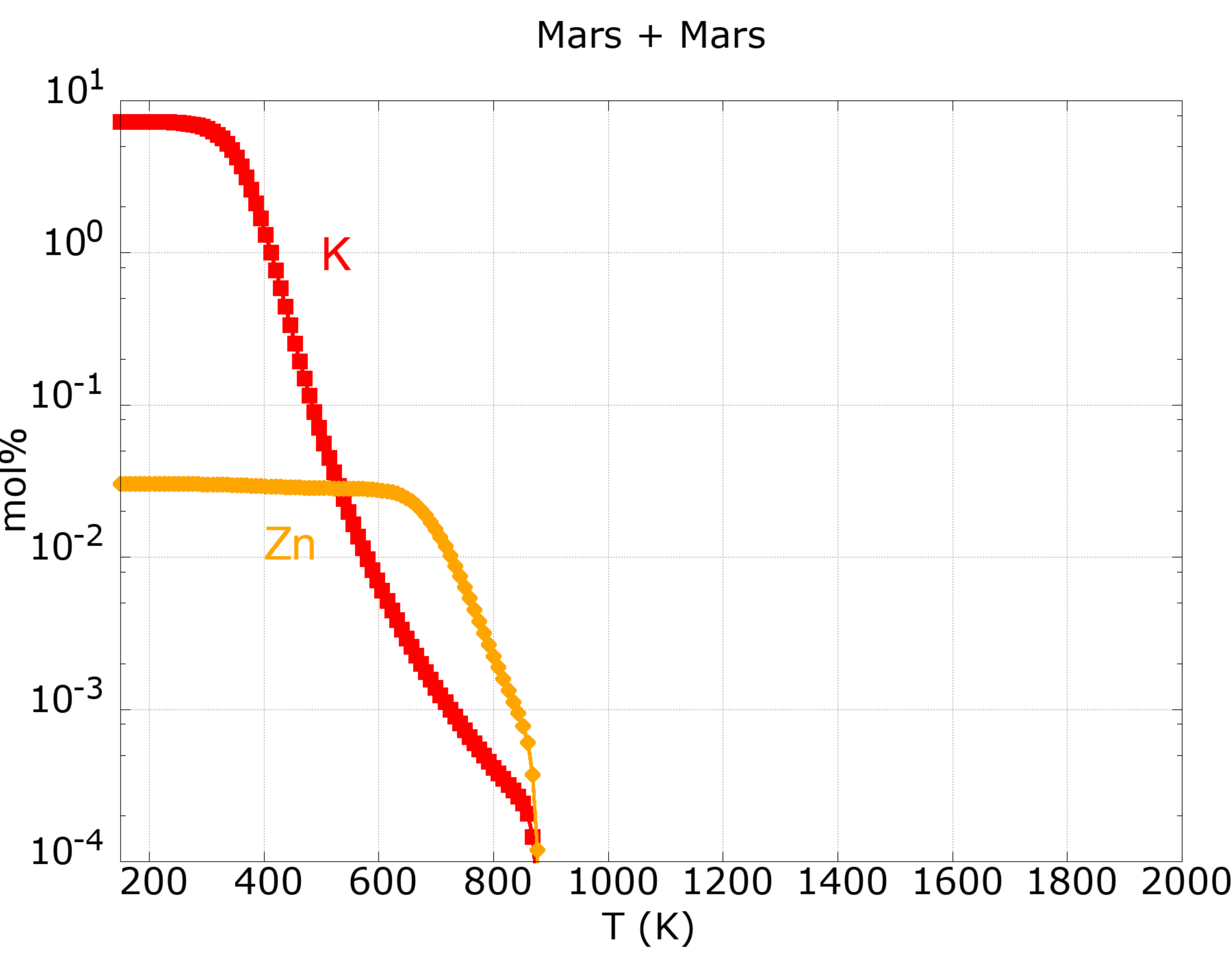}}
{\includegraphics[width=0.5\columnwidth]{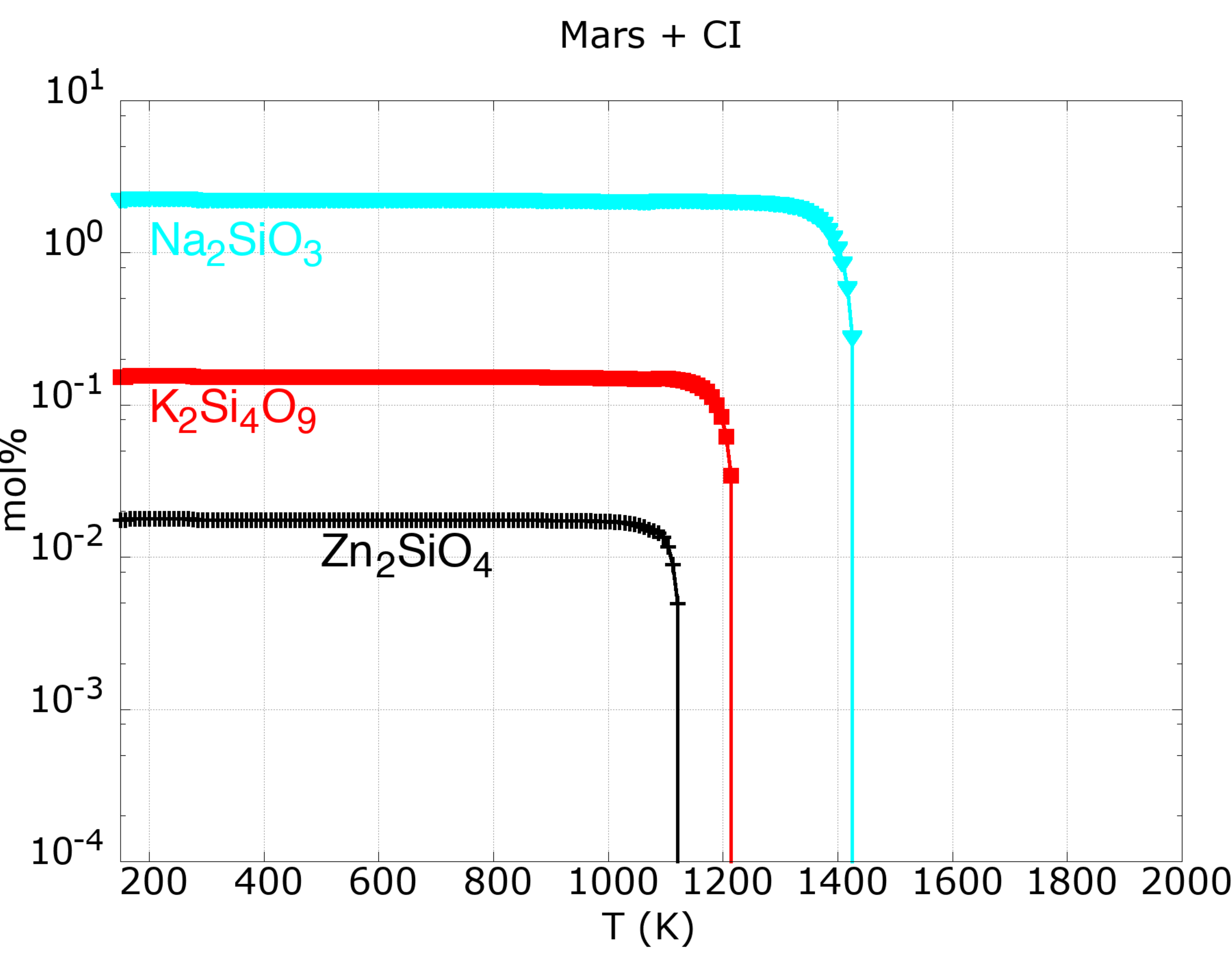}} \\
{\includegraphics[width=0.5\columnwidth]{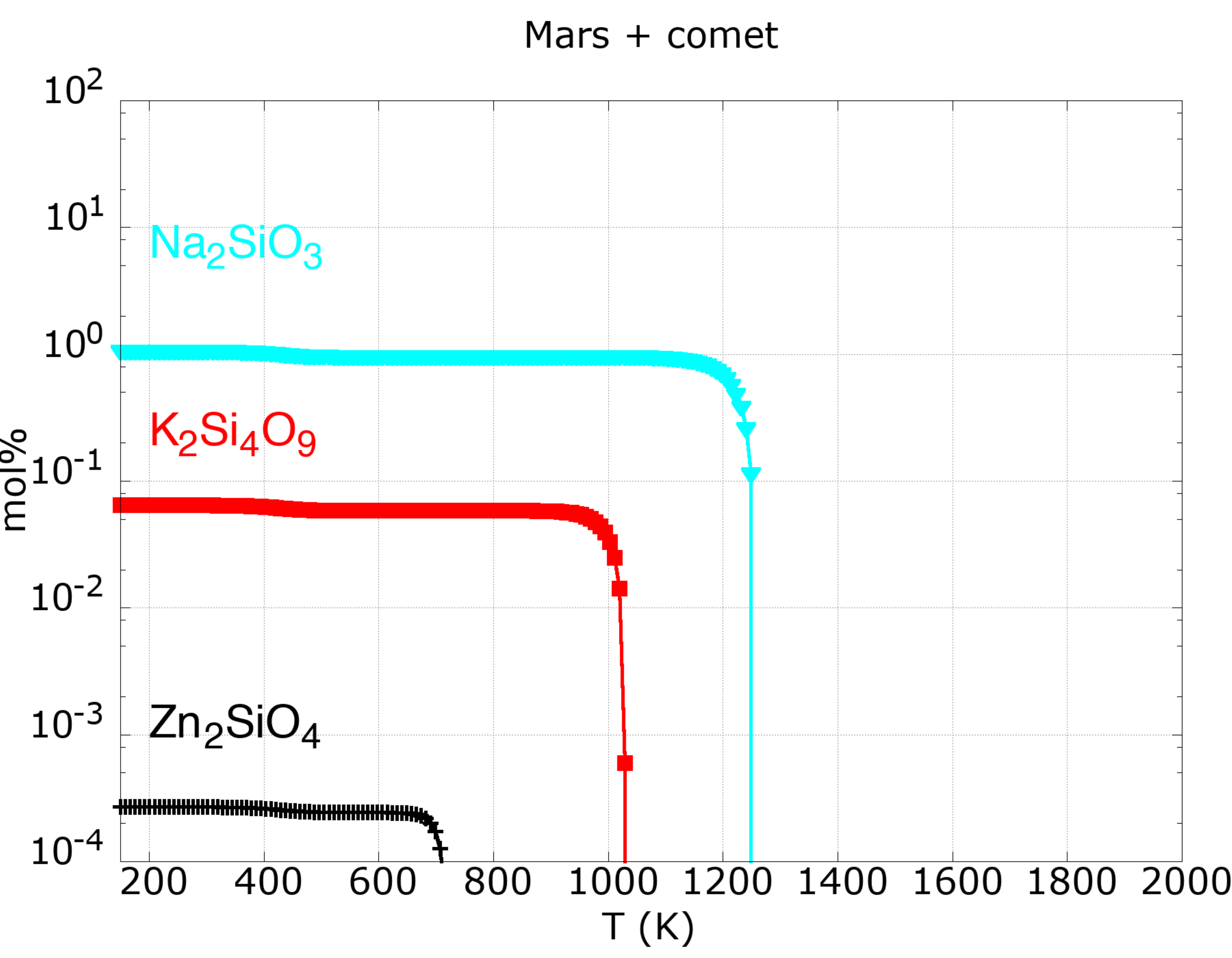}}
\caption{Condensation sequences for K, Zn, and Na compounds (mol\%) that result from the gas mixtures in Table~\ref{table3}. Sodium compounds for Mars+Mars were included in Fig.~\ref{fig2} as, in that case, they represent major species. Note the changes of scales in the y-axis.}
\label{fig3}
\end{figure}

Figure~\ref{fig3} shows the condensation sequence for the more volatiles species. All the considered cases return a very similar behaviour as these volatiles are less effected by the changes of other elemental ratios. Na(g) has higher condensation temperature than K(g) and Z(g) is the last one to condense. \ce{Na2SiO3}, \ce{K2Si4O9} and \ce{Zn2SiO4} are the main respective condensates, together with  Zn in case of Mars+CV, Mars+EH, and Zn and K for Mars+Mars.

\subsection{Solids from cooling melts}
\label{CIPWmelts}

\begin{table*}												
\centering												
\caption{Resulting CIPW-norm of the melt phase. Calculations for the BSM are also performed to compare the resulting CIWP-norm with values  derived by \citet{2016ApJ...828..109R}.}												
\begin{tabular} {l c c c c c c c}												
\hline												
	&+Mars	&	+CV	&	+EH	&	+CI	&	+COMET	&	BSM	& BSM \citep{2016ApJ...828..109R}\\
\hline
Anorthite	&8.08	&	9.69	&	8.24	&	8.35	&	1.39	&	3.16	&  \\
Diopside	&3.08	&	3.63	&	2.97	&	3.13	&	0.00	&	6.89	& 6.97 \\
Pyroxene &	43.41	&	55.48	&	57.58	&	52.35	&	54.97	&	21.03&	21.29\\
Albite&	0.00	&	0.00	&	0.00	&	0.00	&	0.00	&	8.29	& \\
Orthoclase&	0.00	&	0.00	&	0.00	&	0.00	&	0.00	&	0.65& 0.66\\
Olivine&	45.19	&	29.68	&	31.05	&	35.98	&	34.66	&	58.50& 59.22	\\
Ilmenite&	0.25	&	1.52	&	0.17&	0.19	&	6.40	&	0.27&0.00\\
Corundum	&0.00	&	0.00	&	0.00	&	0.00	&	2.59	&	0.00	& 0.00\\
Anorth+Alb& 	0.00	&	0.00	&	0.00	&	0.00	&	0.00	&	(11.45)	& 11.59\\
\hline
Oli/Pyr & 1.04 & 0.53 & 0.62 & 0.68 & 0.63 & 2.78 & 2.78\\
\end{tabular}												
\label{table5}												
\end{table*}

Table~\ref{table5} reports the resulting CIPW-norm if complete equilibration between the melt belonging to Mars and to the impactor occurs (see Table~\ref{table4}). To establish the reliability of our  CIPW-algorithm, we performed calculation using the BSM and compare our result with that in \citet{2016ApJ...828..109R} finding a very good agreement (see the last two columns in Table~\ref{table5}).

The resulting solids  will be generally characterized by pyroxene\footnote{hypersthene in \citet{2016ApJ...828..109R}.}  and olivine, with the former in larger abundances, except for the Mars+Mars case for which olivine is slightly more abundant. Although  our selected  impactors have initially different chemical composition, the resulting CIPW-norm is quite similar for all cases. It is interesting to note that diposide  (\ce{CaMgSi2O6}) is not predicted for a cometary impactor, while corundum (\ce{Al2O3}) is a tracer of that impact. Enstatite and forsterite will be largely stable and common compounds for all the considered cases. Albite (\ce{ NaAlSi3O8}) and orthoclase (\ce{ KAlSi3O8}) are not present in the solids  because Na and K are totally vaporized after the impact (see Tables~\ref{table3} and~\ref{table4}).

\section{Discussion}
\label{discussion}

\subsection{Dust composition}
\label{outlook}

Our calculations clearly show different  behaviour when compared with the classical condensation sequence with a solar composition (see Fig.~\ref{fig5}). One of the main reasons is the amount of H, C and O in our systems that is  very different from the solar values. Moreover, in our calculations the amount of Fe, Mg, and Si is of the same order of magnitude  as O. This is not the case for the Solar Nebula where H is predominant, C is comparable with O and Fe, Mg, Si are orders of magnitude smaller than O \citep{2009ARA&A..47..481A}. Here we try to qualitatively understand our results and emphasize the differences from the well known condensation sequence of the Solar Nebula.

 The stability of forsterite (\ce{Mg2SiO4}) and enstatite (\ce{MgSiO3}) is driven by the Mg/Si ratio  where higher Mg/Si  ratios ($>1$) favour forsterite while lower Mg/Si ratios ($<1$) favour enstatite \citep{2001A&A...371..133F}. However, at very high temperature, forsterite can still be the first magnesium silicates that condenses out before being converted in enstatite \citep{2001A&A...371..133F}. From Table~\ref{table3} we see that the Mg/Si is well below 1 in all cases and, as expected the dust is generally enstatite-rich (see Fig.~\ref{fig2}). The excess of Si that is not consumed in the magnesio-silicates will then be bound with O to form stable \ce{SiO2}. Generally \ce{SiO2} tends to be more stable than iron-oxides (see for example ellingham diagrams in \citet{DeHoff1993}). Fe and \ce{SiO2} can then react to form iron-rich silicates.  If oxygen is still available for reaction it will start to bind iron to form iron-oxides. If there is lack of oxygen, iron will be mainly in the metallic form. The presence of sulfur further modifies the expected composition as sulfidation of Fe occurs. 

As a consequence, looking at the elemental ratios reported in Table~\ref{table3} the behaviours found in Fig.~\ref{fig2} become clearer. Let us consider the two extreme cases of the Fe/O ratio, Mars+comet (Fe/O=0.03) and Mars+CV (Fe/O=1.19). In the case of Mars+comet we have Mg/Si=0.55. As such we expect the oxygen to form mainly enstatite \ce{MgSiO3}. Then we expect the appearance of \ce{SiO2} as there is Si in excess with Si more abundant than Fe (Fe/Si=0.44). As  Fe/O=0.03 and  (Fe+Si)/O=0.09, there is a large reservoir of oxygen to oxidise the iron which, indeed, is found in ferrosilite \ce{FeSiO3}. 

On the other hand, let us focus to the Mars+CV case. Here the small amount of Mg will form magnesium-silicates, then the excess of Si will form \ce{SiO2} and \ce{FeSiO3}. The Fe/Si=3.43 tells us that there is iron in excess compared to Si.  The Fe/O=1.19 and (Fe+Si)/O=1.53 also tell us that there is not enough oxygen available to oxidise all the iron. As a consequence, we expect to see a certain amount of iron to be stable in its  metallic form, and \ce{FeS} given the presence of sulfure in the mixture. Sulfur is also present in the case for Mars+EH, Mars+CI and Mars+Comet. If large amount of sulfur is available, as in the  the Mars+EH impact, MgS  also becomes a stable sulfide.

Interesting cases are also the Mars+Mars and Mars+CI. Here, the high Fe/Si ratios but low Fe/O ratios return large amount of several iron oxides: there is no enough Si to form large amount of iron-silicates, thus the O binds directly the iron in several iron-oxides. The Mars+EH impact clearly shows the effect of the sulfide-rich impactor given the  presence of MgS together with FeS.

Our calculations show that different impactors result in dust with traceable different composition. This open the possibility to identify the individual composition of the impactor from the determination of the dust composition of Phobos and from the samples collected by the MMX mission.

In conclusion, dust from different bodies will be characterized by different i) degrees of iron-oxidation, ii) presence of iron-silicates and/or iron-oxides, iii) amount of sulfides, iv) amount of carbon and ice. All the condensation sequences return  a generally poor content of olivine and pyroxene, with a preference of the latter. A qualitative analysis of different elemental ratios can then be useful to derive the chemistry of impactors that are not considered in this work.

\subsubsection{Carbon, water ice and other volatiles}
\label{carbonwater}

In our calculations we  see the appearance of solid C in the case of Mars+CV, Mars+CI and Mars+comet, while \ce{SiC} is the most stable C-bearing compound in the case of Mars+EH . Mars+CV, Mars+CI and Mars+comet have carbon and hydrogen in the gas mixture, while in the Mars+EH case there is carbon only.

The carbon-to-oxygen ratio (C/O) is an important parameter that determines the presence of solid carbon,  water vapour and other oxides. At high temperature CO(g) will consume all the C available before allowing the formation of water vapour \citep{1975GeCoA..39..389L}. This has strong implication in the formation of complex organics and water (if hydrogen is present in the system).

The chemistry of carbon cannot be totally determined by thermodynamic equilibrium. The behaviour of CO(g) in a \ce{H2O}(g)-\ce{H2}(g) gas in the temperature range of  $100< T\rm{(K)}<700$~K is ruled by kinetics and the environmental condition. The classical transition around $T\sim700$~K where \ce{CO}(g) is transformed in \ce{CH4} can be described, for example, by a reaction of the type $n\ce{CO}+(1+2n)\ce{H2} \rightleftharpoons \ce{C_{n}H_{2n+2}} + n\ce{H2O}$. This can be noticed in Fig.\ref{fig5}(right)  where $n=1$ and the reaction is  $\ce{CO}(g)+3\ce{H2}(g)\rightleftharpoons \ce{CH4}(g) + \ce{H2O}(g)$.

In fact, this reaction is just a first simplified transcription of the many Fischer-Tropsh-like reactions  that can occur in this temperature range and at different \ce{H2}/\ce{CO} ratios \citep{2006M&PS...41..715S,2008ApJ...673L.225N}. Fischer-Tropsh processes produce complex organics  on the surface of dusty grains in the presence of the right catalyst. There are different catalysts with their own properties, but usually, in astrophysics context, the iron-based Fischer-Tropsh is the most considered because the abundances of this element in the Solar Nebula \citep{1988oss..conf...51F,2009ARA&A..47..481A}.
 
 There are numerous competitive reactions that determine the rate and the production of organics \citep{1988oss..conf...51F} and several theoretical models have been largely described in material science (see for example the work of \citet{Zimmerman1990}). However, the resulting amount of organics   is extremely difficult to calculate theoretically when an astrophysical environment is considered. This comes from the large uncertainty in determining, for example, the amount and the surface of catalyst available.
  
 \begin{figure}
\center
{\includegraphics[width=0.5\columnwidth]{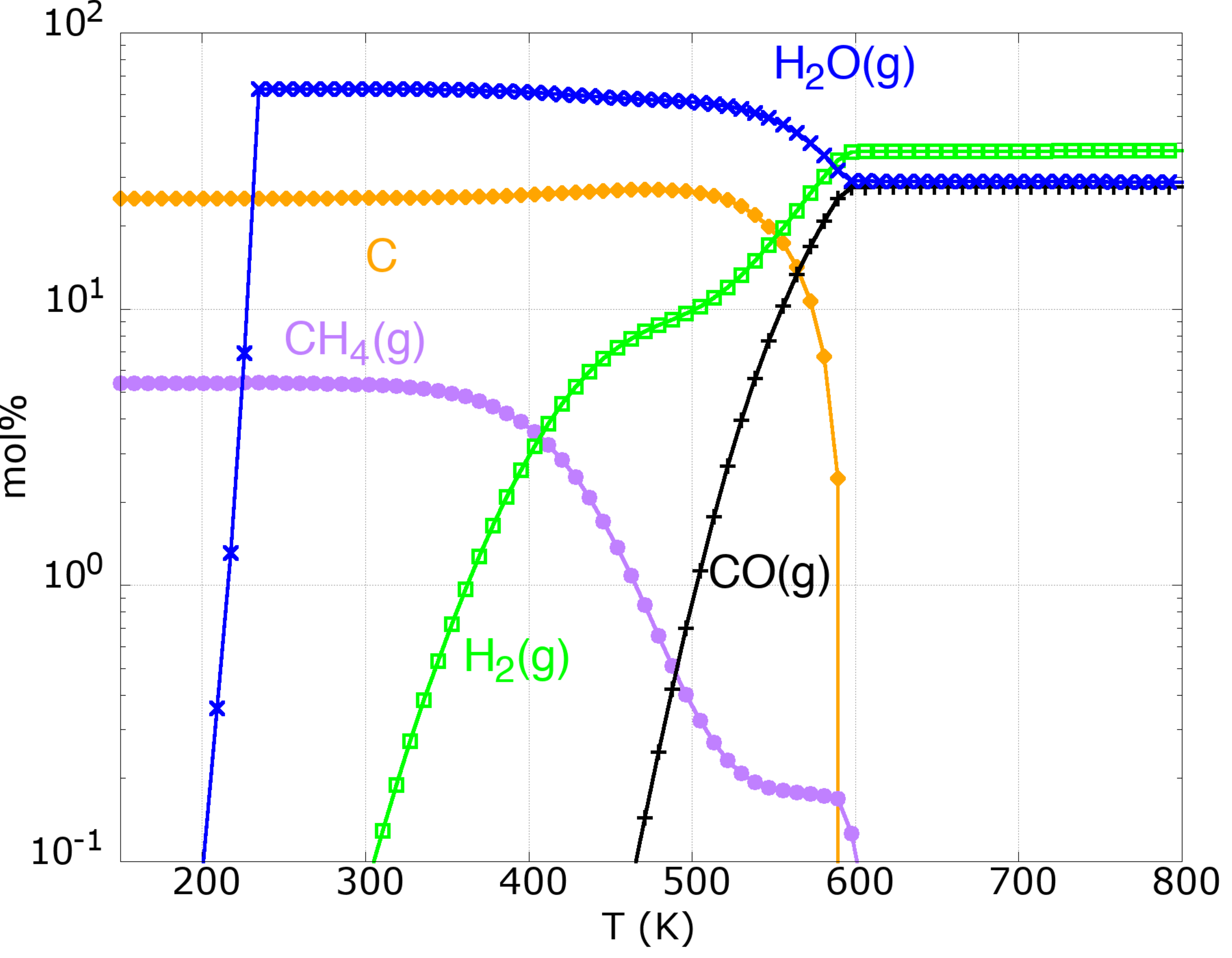}}
\caption{Thermodynamic calculations in the +comet case shows a possible pathway for the condensation of graphite. In fact, together with the well known transformation of \ce{CO}(g) in \ce{CH4}(g) ($\ce{CO}(g)+3\ce{H2}(g)\rightleftharpoons \ce{CH4}(g) + \ce{H2O}(g)$), we do see the following reaction: $\ce{CO}(g)+\ce{H2}(g)\rightleftharpoons \ce{C} + \ce{H2O}(g)$  that occurs at temperature lower than T=600~K.}
\label{fig4}
\end{figure}
Nevertheless,  the possible pathways to the formation  of carbon-rich material are vastly more numerous. In our Mars+comet case, for example,  the reaction $\ce{CO}(g)+\ce{H2}(g)\rightleftharpoons \ce{C} + \ce{H2O}(g)$ becomes active as well. This is clearly seen in figure~\ref{fig4} where \ce{H2}(g) and \ce{CO{g}} are depleting as \ce{CH4}(g), \ce{H2O}(g) and C become more stable. In this work we do not perform any kinetics calculation and, since we have only few carbon-solids in our list, we can only suggest that the presence of C and H in Mars+CI, Mars+Comet and Mars+CV impacts can produce complex organics and carbon-enriched dust. The case of Mars+CV is then extremely interesting  because there may be enough metallic iron, together with carbon, to enhance the production of complex organics.
  
Equilibrium calculations return an efficient evaporation of the carbon rich dust present in the impactor. However, the rate of vaporisation  will be driven by the physical and chemical properties of the carbon species. For example, carbon rich insoluble organic material (IOM), if present,  could survive the impact as it is refractory \citep{2006mess.book..625P}. This could reduce the amount of carbon released in the gas phase. Nevertheless, the presence  of carbon in the MMX samples in the form of new condensates and/or in IOM will still be tracer of the nature of the impactor. 
 
In section~\ref{results} we pointed out that the gas-mixture is volatile-enriched. Figure~\ref{fig3} shows that K-, Na-, and Zn-silicates are stable as the temperature drops. Mono atomic Zn is also predicted for the Mars+EH and Mars+Mars case. There are no dramatic differences between the considered impactors  when K, Na and Zn are taken into account. 

In conclusion, the condensing dust will be volatile (and in some cases, also carbon) enriched compared to the solids  that result from the cooling melts. The presence of volatile-rich dust in MMX samples will thus indicate that vaporization followed by condensation had occurred and no volatiles left the system. Moreover, since Na, K and Zn condense at different temperature, their presence/absence could return information on the temperature at which  aggregation of a given sample  occurred.

Mars+CI and Mars+comet are the only cases for which, at low temperature, we  see condensation of water vapour into ice. The presence of ice will favour secondary alteration of the dust allowing, for example, the formation of phyllosilicates \citep{1998M&PS...33.1113B}.

\subsection{Solids composition}
\label{melts}

In table~\ref{table5} we reported the resulting CIPW-norm of the solids if complete equilibration between Mars and impactor occurred. The composition of the solids generally comprises  olivine (forsterite and fayalite) and pyroxene (enstatite and ferrosilite). \citet{2016ApJ...828..109R} calculated the CIPW normative mineralogy for a Mars, Moon and IDP like impactor. They found that the resulting composition of a mars-like impactor would be olivine and pyroxene rich. In particular their CIPW-norm for Mars is characterized by high olivine content ($olivine/pyroxene > 1$). 

Our results for the Mars+Mars case show that the solids (deprived of all the vaporized material) will have an $olivine/pyroxene \sim 1$, whereas the other cases return a $olivine/pyroxene < 1$. 

There are no dramatic differences between the solids that results from different impactors (except for the aforementioned corundum in the comet case). It is interesting to note that solids that result form cooling melts do not show as much as variation in their composition compared to the dust. The resulting composition of solids appears, thus, to contain less information on the origin of the impactor compared to the large quantities of clues that can be derived from condensed dust. Nonetheless, the composition of  melts can be affected by different cooling conditions, microgravity and gas fugacities. \citet{NAGASHIMA2006193} and \citet{2008070620c} performed laboratory experiments on  cooling  forsterite and enstatite melts. They found that different cooling rates and microgravity can alter and even suppress crystallisation only allowing the formation of glass material. Further experimental investigations  are already planned in order to derive  predictions of the composition of the cooling melts.

On the physical point of view, condensates and solids from melts may be distinguished by their different crystalline structure, microporosity, zoning, interconnections between different phases. Indeed, the resulting physical properties of dust from gas and solids from melts are determined by many factors \citep{nishinaga2014handbook}. Furthermore \citet{2017ApJ...845..125H}  showed that while the size of dust would be in the order of 0.1$\sim$10$\mu$m, solids from melts can reach 1$\sim$10m in size and then they can be grinded down to $\sim$100$\mu$m. We could, thus, expect to find different size distributions when dust and solids are compared.

\subsection{Infrared spectra of Phobos}
\label{infraredspectra}

\citet{2011P&SS...59.1308G} presented a detailed investigation on the possible composition of the dust and rocks present on the Phobos surface. They suggested that the ``blue'' part of Phobos is consistent with phyllosilicates while the ``red'' region is compatible with the presence of feldspar. No bulk chondrite compositions are able to reproduce the current observation \citep{2011P&SS...59.1308G}.

Phyllosilicates are not product of condensations but derive from secondary alterations of silicates \citep{1998M&PS...33.1113B} and, as a consequence, they are not predictable with our calculations, although we do have all the dust (silicates) at the base of their formation. The major feldspar compounds are orthoclase (\ce{KAlSi3O8}), albite (\ce{NaAlSi3O8}), and anorthite (\ce{CaAl2Si2O8}). Not all of them are compatible with our model as Na and K are separated from Al after the impact and others are the predicted stable compounds. On the other hand we do see the formation of  anorthite (see Table~\ref{table5}).

Nevertheless \citet{2011P&SS...59.1308G}  stressed that more fine modelling is needed as a  mixture of different materials made of fine grains could also produce the observed trends. This becomes important as previous modelling   focused on the analysis of the spectral properties of ``external'' objects (such as different types of asteroids) to match the observed spectra, as the capture scenario suggests. The impact-generated scenario imposes to re-think this approach.

\citet{2016ApJ...828..109R} analysed the resulting composition of the melts generated by different impactors in order to find a match with the observations. They concluded that more than the  melt, the gas-to-dust condensation in the outer part of an impact-generated disk could be able to explain the Phobos  and Deimos spectral properties. This further suggests that our derived dust  may play an important part in producing the observed trends. Moreover, since the \citet{2017ApJ...845..125H} disc model shows that melt would be mixed with the gas, the combined effect of {\it dust} and {\it solids} should be taken into account.

In this section we try to predict the effects of our mixed material (dust plus solids)  on the infrared spectra. In the introduction we presented possible mechanisms  listed by \citet{2011P&SS...59.1308G} that could be able to reproduce the observed trend in the VIS-NIR. Here we recall them and compare with our results:

i) {\it Low percentage of iron-rich olivine and pyroxene can reduce the spectra}. Our resulting dust mixtures generally have a very low concentration ($\sim 1/$~mol\% ) of iron-rich olivine (fayalite, \ce{Fe2SiO4}) and pyroxene (ferrosilite, \ce{FeSiO3}). Only  the high temperature region of Mars+CI shows a larger amount of fayalite (see Fig.~\ref{fig2}). 

ii) {\it A mixture of opaque material (metal iron, iron-oxide, and carbon) reduce the emissivity}. We do have a metallic iron-rich dust that  results from several impactors (Mars+CV, Mars+EH, and Mars+Mars only at low temperature).  Carbon dust is also seen in our calculations  (Mars+CV, Mars+CI, Mars+comet). Moreover, together with Fe, iron sulfide (\ce{FeS}),  that we see in Mars+CV, Mars+EH, Mars+CI, Mars+comet, is opaque and featureless in the NIR , but may be recognizable in the MID-IR \citep{Wooden2008,2011ppcd.book..114H}.

iii) {\it  Quenched material lacks of perfect crystalline structure and, thus, reflectance}. Solids from the melts, in our final assemblage, can have the characteristics suggested by \citet{2011P&SS...59.1308G}.

iv)  {\it  The reflectance of fine grains is reduced}. The average size of the condensed dust is in the order of  0.1$\sim$10$\mu$m \citet{2017ApJ...845..125H}.

Our proposed model of Phobos as a result of  accretion of dust from gas condensation and solids from melts, together with our derived chemical composition looks promising when discussing spectra. Nevertheless there are some aspects that needs further investigation: i) it is important, at this point, to derive the MIDIR spectra of our propose mixtures, and then ii) estimate the effect of space weathering on it and  on the resulting albedo. These points can be set as main topic for future works.

 \subsection{Limitations}
 \label{limitations}
In this work we assume thermodynamic equilibrium (where all the reactions rates are much shorter than the disk cooling timescale) and mass conservation. All the material is  available for reaction until the equilibrium is reached at any given temperature. This may be not always the case. As dust condenses out from the gas, it can be subject to different drag forces and it can separate from the current environment. This process can lead to the so called ``fractionated'' condensation sequence. If this is the case, a given dust grain can become representative of the temperature, pressure, gas mixture and dust species that were present when its condensation occurred. 

Moreover, dust from secondary condensations   (from a now fractionated gas) may then form. These fractionated and incomplete condensation sequences have been the subject of several studies  \citep{2000M&PS...35..601H,2009M&PS...44..531B,2016MNRAS.457.1359P} which  all show that different pathways of condensation can depart from the main line. For example, \citet{2016MNRAS.457.1359P} showed that, starting with a gas of solar composition, a mixture of  enstatite-rich  and \ce{SiO2}-rich dust can be produced in case of systematic sharp separation between dust and gas. \ce{SiO2} is a condensate that is not predicted (``incompatible'') when solar abundances are considered. 

In this work we do not perform fractionated condensation sequences as numerous are the possible pathways. However, in the same way, the presence of some ``incompatible'' condensates together with ``predictable'' dust in the same MMX sample may point out to incomplete or secondary condensation.

As mentioned,  \citet{2017ApJ...845..125H} demonstrated that there will be likely no complete equilibration between the melts of Mars and the melts of the impactor. What is likely to occur is a wide spectrum with different degrees of equilibration. A random sampling of solids  may, thus, show material that come from Mars, the impactor or several degrees of mixing.

In our calculations we kept the pressure constant and fixed at $P=10^{-4}$~bar as in \citet{2017ApJ...845..125H}. In general, lower pressures (in order of magnitude)  move the condensation of the dust toward lower temperatures \citep{Yoneda1995,1998A&A...332.1099G}. In this case, for our given temperature, more material could vaporize and go into the gas-phase. Increasing the pressure (in order of magnitude) has the opposite effect as the condensation temperature increases. As a consequence, we could observe different amount of Fe, Mg, and Si moving to the gas phase.  Changes in disk pressure may also occur if  large amount of volatiles are injected in the system after the impact. This could be the case of the Mars+comet impact where the release of \ce{H2O}(g) and \ce{CO(g)} could change the total pressure in the disk increasing it, or when a water-rich Mars is considered \citep{2017ApJ...845..125H}.  Observed deviations from the predicted trends could then be associated to strong variation in the pressure in the Mars' Moons formation region of the disk or, in fact, to a radial gradient of temperature and pressure in the disk. As reference for future experimental work we report in appendix B (see Fig.\ref{fig6}) the partial pressures of the major gas component for Mars+CI, Mars+CV, and Mars+comet impacts. These are the impacts that produce the larger amount of gas such as \ce{H2O}(g) and \ce{CO}(g). These values can be used to set up the  conditions in which experiments can be performed.

\section{Conclusions}
\label{conclusions}

In this work we used thermodynamic equilibrium calculation to investigate the chemical composition of dust (from condensing gas) and solids (from cooling melt) as the building blocks of Phobos and Deimos in the impact-generated scenario with the thermodynamic conditions of \citet{2017ApJ...845..125H}.

We found that  dust and solids have different chemical and physical properties. Dust carries more information on the impactor than the solids. Our results show that it would be possible to distinguish from different types of impactors as each case returns several unique tracers in the dust: a Mars+CV  has large quantities of metallic iron, \ce{SiO2}, iron sulfides and carbon; Mars+Comet has pyroxenes and the largest carbon and ice reservoir. Mars+EH impact has dust with high metallic iron content, \ce{SiO2}, sulfides (\ce{FeS} and \ce{MgS}) and traces of \ce{SiC}. Impact with Mars-like objects returns several iron-oxides, and the dust in Mars+CI has iron-oxides, water ice and carbon.

The presence/absence of metallic iron, iron-silicates, iron-oxides, sulfides, carbon and water ice can be considered  as clues of different impactors. Deviations from the derived compositions can be then ascribed to fractionated condensation sequences and/or strong variations in the disk pressure  and/or impactors with different elemental composition than investigated in this study.

The giant impact scenario imposes to re-think  the dust modelling for the infrared spectra, as Phobos, in this case, would be made of a complex mixture of dust and solids and not of a pre-built object as the capture scenario suggests. A  qualitative analysis suggests that our derived composition of dust and solid can be compatible with the characteristic of the Phobos VIS-NIR spectra. 

In conclusion, the proposed scenario of Phobos  as the result of accretion of dust and solid in an impact-generated disk can reconcile with both the dynamical and spectral properties of the Mars' moon. Our dust tracers can be then used in the analysis of the samples returned by the JAXA's MMX mission.

\software{HSC (v8; \citet{roine2002outokumpu})}

\acknowledgments

The authors wish to thank the anonymous referee for their comments and suggestions that let us investigate our assumptions in more details which improved the manuscript. The authors wish to acknowledge the financial support of ANR-15-CE31-0004-1 (ANR CRADLE), INFINITI (INterFaces Interdisciplinaires Num\`erIque et Th\`eorIque), UnivEarthS Labex program at Sorbonne Paris Cit´e (ANR-10-
LABX-0023 and ANR-11-IDEX-0005-02). PR has been financially supported, for his preliminary contribution to this work, by the Belgian PRODEX programme managed by the European Space Agency in collaboration with the Belgian Federal Science Policy Office. SC, RH, and HG acknowledge the financial support of the JSPS-MAEDI bilateral joint research project (SAKURA program). HG also acknowledges JSPS KAKENHI Grant Nos. JP17H02990 and JP17H06457, and thanks the Astrobiolgy Center of the National Institutes of Natural Sciences (NINS).

%

\vspace{5mm}

\appendix
\section{Solar condensation sequence}
\label{solar}
Condensation sequences calculated with solar abundances are very distinctive and characterized by smooth transitions between refractories, main silicates and metallic iron and low temperature material (such as sulfides) \citep[]{1979Icar...40..446L,Yoneda1995,2003ApJ...591.1220L,2011MNRAS.414.2386P}.

\begin{figure}
\center
{\includegraphics[width=0.4\columnwidth]{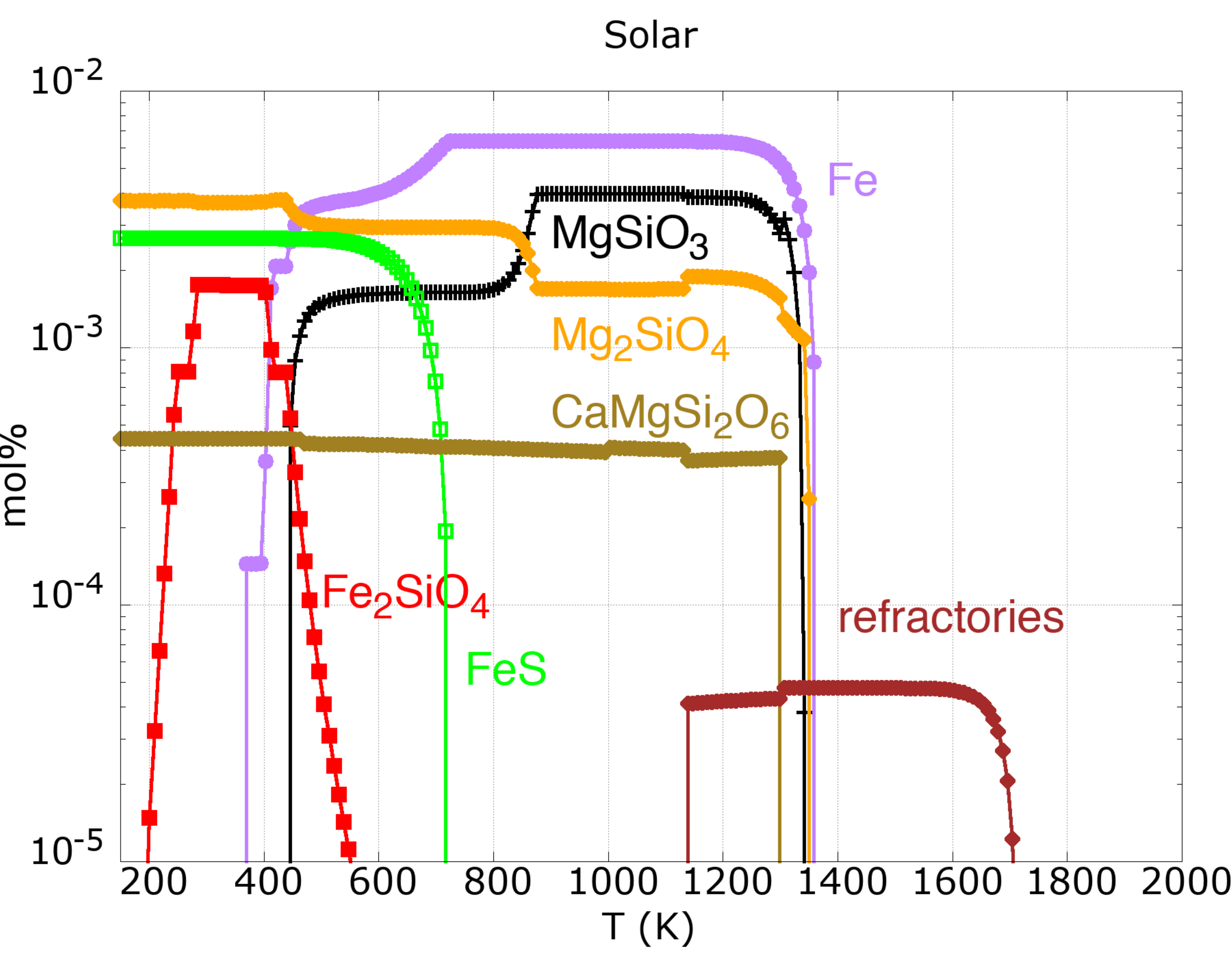}}
{\includegraphics[width=0.4\columnwidth]{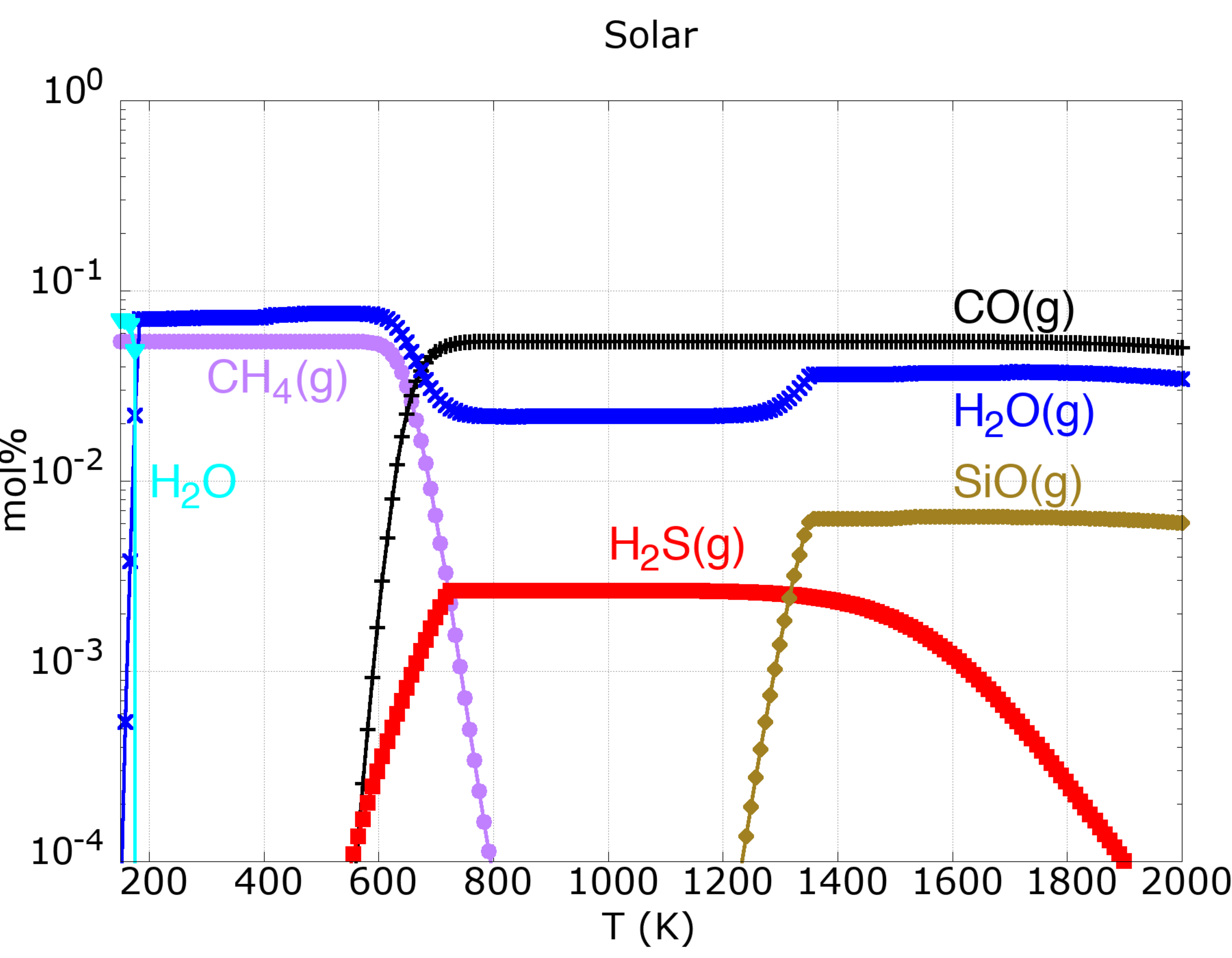}}
\caption{Condensation sequence for major solids (left) and gas (right) starting with an initial gas with solar composition. There is a very good agreement between these results and those reported in \citet{2011MNRAS.414.2386P} and reference therein.}
\label{fig5}
\end{figure}

Fig.~\ref{fig5} reports the condensation sequence calculated with our thermodynamic system and the solar abundances listed in Table~\ref{table1} in the temperature range of $150< T\rm{(K)}<2000$  and pressure $P=10^{-4}$~bar. The small fraction of solids that are stable at high temperature ($T>1400$~K) are the refractories calcium-alluminum silicates while at $T\sim1400$~K iron, forsterite and enstatite condense. At lower temperature we see the formation of troilite (\ce{FeS}) and fayalite (\ce{Fe2SiO4}). \ce{CO}(g), \ce{H2O}(g) and \ce{SiO}(g) are the main O-binding gaseous species. At $T\sim 700$~K we see the conversion of \ce{CO}(g) to \ce{CH4}, while \ce{H2S} is the main sulfur-binding gas until the condensation of troilite.

 These results are in very good agreement with  previous works \citep{1979Icar...40..446L,Yoneda1995,2003ApJ...591.1220L} on the solar condensation sequence and make us confident on our built system.

\section{Partial pressures of main gaseous compounds}
\label{partialpressure}

\begin{figure}
\center
{\includegraphics[width=0.4\columnwidth]{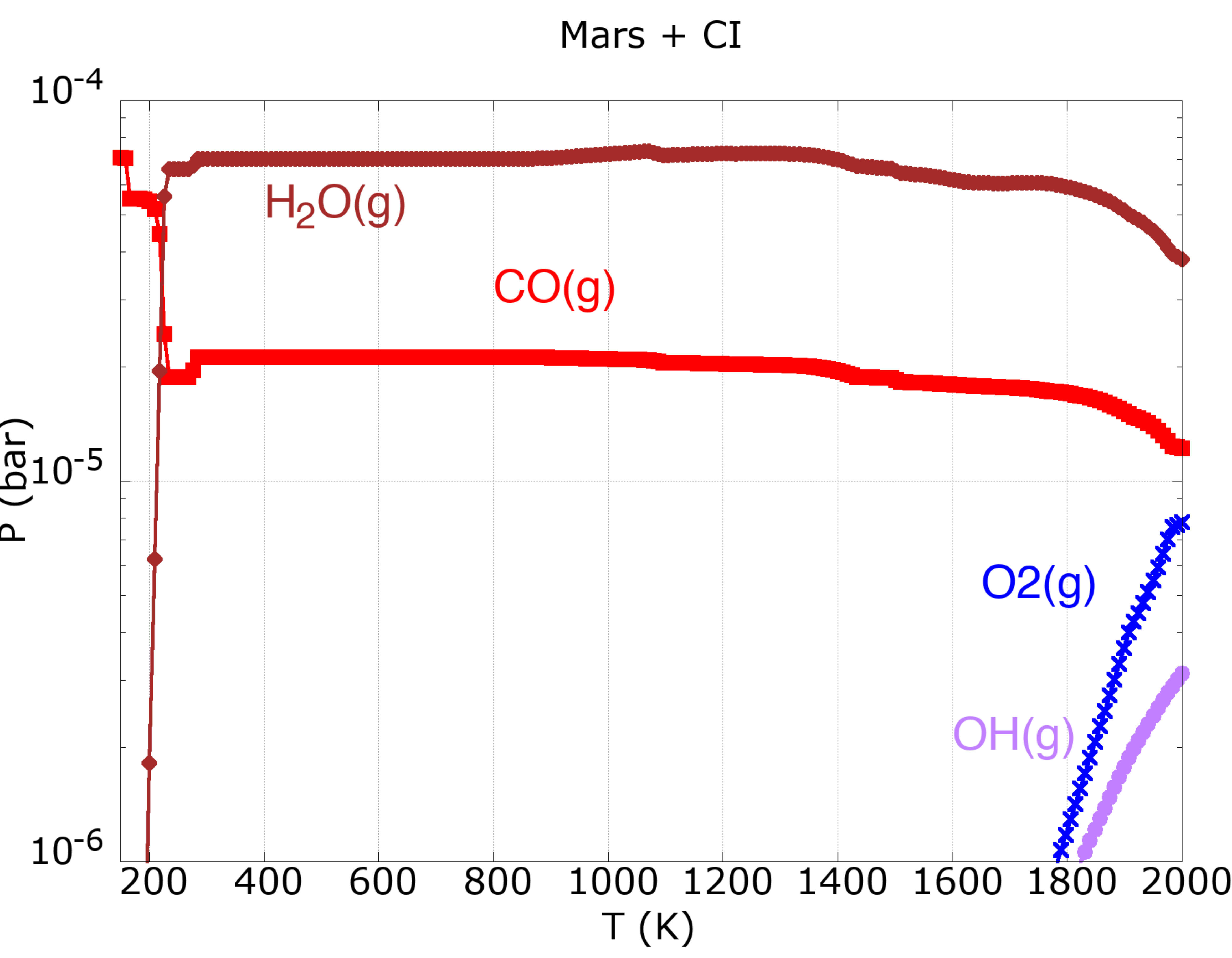}}
{\includegraphics[width=0.4\columnwidth]{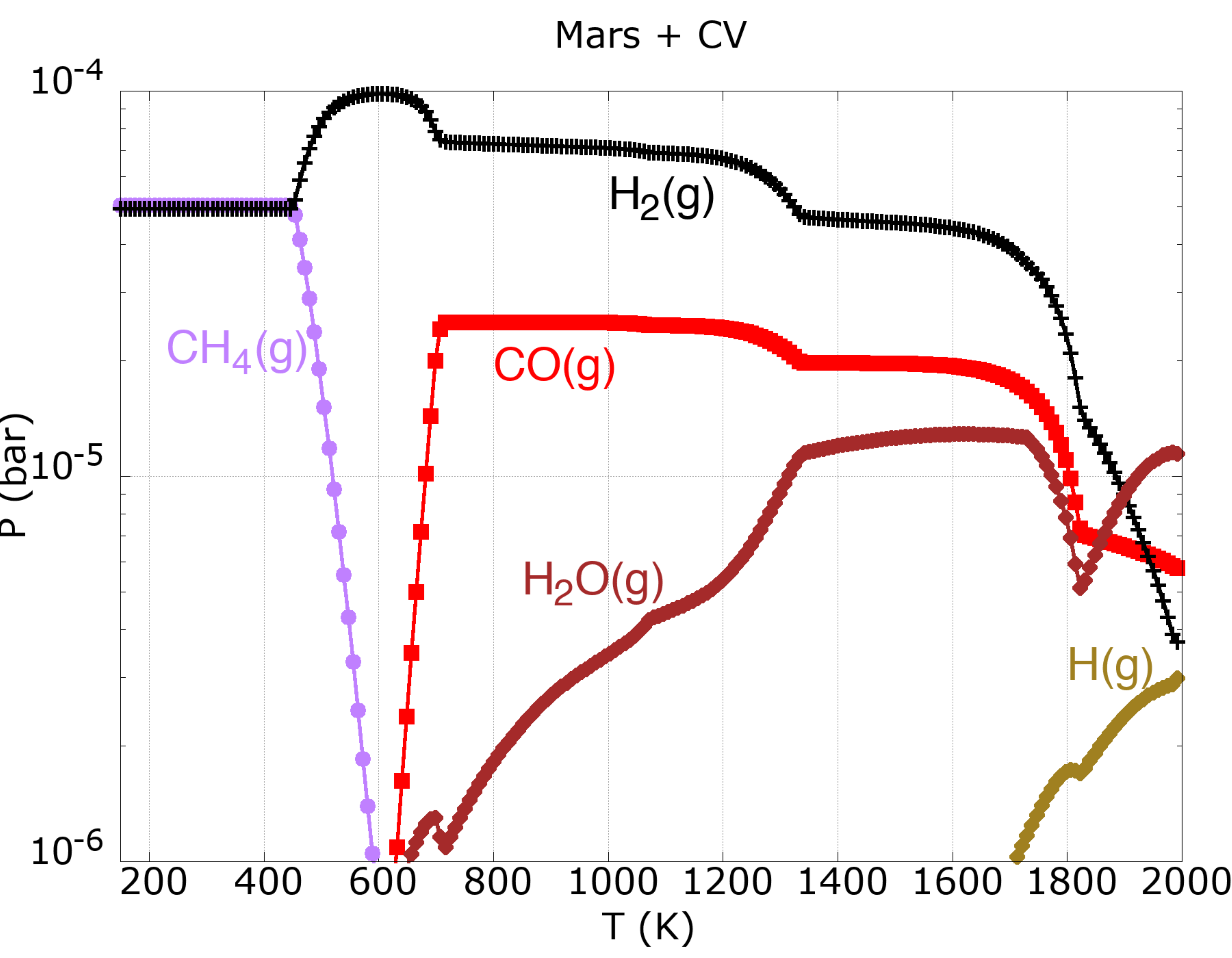}}
{\includegraphics[width=0.4\columnwidth]{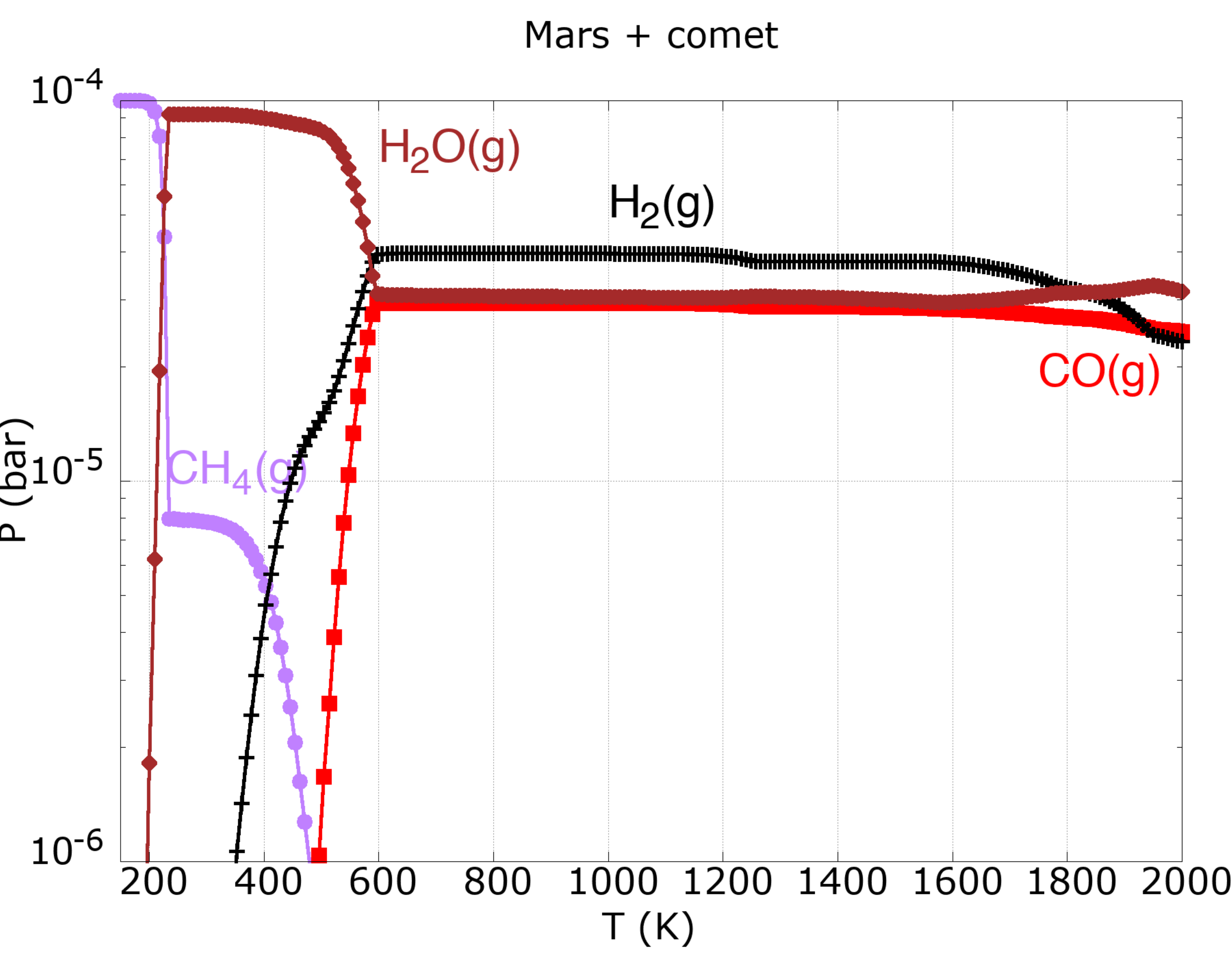}}
\caption{Partial pressure for major gas in the Mars+CI, Mars+CV and Mars+comet impacts. The total pressure of the system is $P=10^{-4}$ bar.}
\label{fig6}
\end{figure}

 In figure~\ref{fig6} we report the partial pressures for the main gaseous species as they result from our equilibrium calculations. The chosen case are Mars+CI, Mars+CV and Mars+comet. These are the impacts that introduce  H, C, and O in the system. In our calculations ideal gas are considered. As a consequence their partial pressure can be taken as a proxy for their fugacity. These values can be used  to set the initial gas environment in further experimental studies.

\newpage
\bibliographystyle{apj}
\bibliography{biblio}



\end{document}